\documentclass{article}

\usepackage{graphicx}%
\usepackage{multirow}%
\usepackage{amsmath,amssymb,amsfonts}%
\usepackage{amsthm}%
\usepackage{mathrsfs}%
\usepackage[title]{appendix}%
\usepackage{xcolor}%
\usepackage{textcomp}%
\usepackage{manyfoot}%
\usepackage{booktabs}%
\usepackage{algorithm}%
\usepackage{algorithmicx}%
\usepackage{algpseudocode}%
\usepackage{listings}%
\usepackage{caption}
\usepackage{lineno}
\usepackage{ulem}
\usepackage[super,sort&compress]{natbib}
\usepackage{float}
\usepackage{url}
\usepackage{hyperref}
\usepackage{bibunits}

\newcommand{\supercite}[1]{\textsuperscript{\cite{#1}}}

\usepackage{setspace}
\title{\textcolor{black}{Imaging exoplanets below the telescope stability limit}}
\author{
  Rodrigo Ferrer-Chavez$^{1,2,*}$ \and
  Jason J. Wang$^{1,2,\dagger}$ \and
  Katherine L. Bouman$^{3,\dagger}$ \and
  William T. Freeman$^{4}$ \and 
  Laurent Pueyo$^{5}$ \and 
  Brandon Y. Feng$^{4}$ \and 
  Marshall D. Perrin$^{5}$
}
\date{}

\begin{document}


\maketitle








$^{1}$Department of Physics and Astronomy, Northwestern University, 2145 Sheridan Rd, Evanston, IL 60208, USA

$^{2}$Center for Interdisciplinary Exploration and Research in Astrophysics (CIERA), Northwestern University, 1800 Sherman Ave, Evanston, IL 60201, USA

$^{3}$Departments of Computing and Mathematical Sciences, Electrical Engineering, and Astronomy, California Institute of Technology, Pasadena, CA 91125 USA

$^{4}$ Computer Science and Artificial Intelligence Laboratory, Massachusetts Institute of Technology, Cambridge, MA 02139 USA

$^{5}$Space Telescope Science Institute, 3700 San Martin Drive, Baltimore, MD 21218, USA

$^{*}$Corresponding author.

$^{\dagger}$These authors contributed equally to this work. 

\begin{bibunit}[sn-nature-custom]
\abstract{\textcolor{black}{Exoplanet imaging is the primary method for detailed characterization of planets orbiting beyond 1 au \supercite{currie2022direct}, but it requires suppressing the bright glare of the star to see the faint planet light. In addition to sophisticated instrumentation\supercite{kenworthy2025high,rigby2023science}, effective image processing is necessary for starlight suppression, which continues to be a significant limitation\supercite{godoy2024escape}. Standard algorithms use reference images for starlight modeling, and thus they are limited by inevitable telescope instabilities that change from reference to science observation.\supercite{telfer2024empirical,perrin2018updated}. Addressing this challenge will increase the number of planets amenable to direct imaging, and will be a major step towards imaging exo-Earths with future observatories \supercite{stark2024paths}. Here we present a new physics-based method for starlight subtraction. Our method combines a multi-step optimization with a flexible physical model to produce a starlight model, allowing modeling and compensation of instrument instabilities. We leverage automatic differentiation \supercite{lecun1988theoretical} to optimize millions of model parameters. We demonstrate our method on observations from the James Webb Space Telescope (JWST) NIRCam \supercite{Gardner2006, Rieke2005,Krist2009} coronagraph instrument. We achieve sensitivity 2-15 times superior to standard techniques, including the first detection of 51 Eridani b \supercite{macintosh2015discovery} at 4.3 $\mu$m. This is the first time that such an approach has demonstrated substantially improved sensitivity on real, on-sky coronagraphic imaging. Our framework can be extended to other current and future ground- and space-based observatories, including the Roman Space Telescope and the Habitable Worlds Observatory, potentially relaxing the unprecedented stability requirements needed to make these a success.}}

\maketitle

\section{Main}\label{main}

In order to \textcolor{black}{detect} any planetary signals \textcolor{black}{via direct imaging}, the glare of the star (or stellar point spread function, PSF) must be modeled and removed from our observations. \textcolor{black}{Since even the brightest young giant planets are $\sim10^5$ times fainter than their host stars in the near infrared, modeling} the stellar PSF to the required precision is difficult. \textcolor{black}{This is} due to its strong dependence on \textcolor{black}{factors such as } stellar properties and instrumental aberrations. Until now, the most successful approaches build the PSF model empirically using a library of reference observations taken at other times or wavelengths, or of other stars (e.g., \supercite{lafreniere2007new,soummer2012detection}).  

However, these empirical approaches are fundamentally limited by \textcolor{black}{various sources of instabilities \textcolor{black}{(Figure \ref{fig:optimization})}. In space-based observatories, these mostly include telescope pointing errors and thermal drifts \supercite{perrin2018updated}}\textcolor{black}{, causing differences between reference and science PSFs and reducing our sensitivity\supercite{perrin2018updated, krist2023end}}. Advances in PSF modeling that do not rely as heavily on observatory stability could have transformative implications for \textcolor{black}{exoplanet imaging, for both current and future observatories. For example, one of the main challenges for future missions like the Habitable Worlds Observatory (HWO), with the ambitious goal of imaging Earth analogs around Sun-like stars ($10^{10}$ brightness contrast), is achieving an extreme and unprecedented stability at the picometer level (100 to 1000 times more stable than JWST\supercite{telfer2024empirical,feinberg2026habitable}).}



 An attractive alternative \textcolor{black}{to starlight subtraction via empirical reference images} is model-based PSF subtraction: directly modeling the physics of the instrument to reconstruct the stellar PSF. This approach has been noted for its potential to relax the stability requirements of the telescope \textcolor{black}{and to reach the fundamental Poisson noise sensitivity limit}  \supercite{stark2024paths}. While promising, \textcolor{black}{previous work investigating} model-based PSF subtraction \textcolor{black}{have noted} formidable \textcolor{black}{challenges} given the complexity of optical systems that prevents accurate modeling \supercite{ygouf2013simultaneous,Xin_2024,Pope_2021,DeFurio2025}.  Even in a stable space-based instrument like JWST/NIRCam, PSF subtraction with synthetic PSFs has been only used in extraordinary circumstances, where \textcolor{black}{empirical reference libraries are exceptionally low quality}, and it has never produced sensitivity superior to standard techniques (e.g.,\supercite{lawson2_024,greenbaum2023first}).

\textcolor{black}{Fortunately, nowadays the computational challenges of model-based PSF subtraction can be mitigated with GPU-accelerated differentiable models. Differentiable models leverage automatic differentiation \supercite{lecun1988theoretical}, or autodiff, to compute analytic derivatives even with millions of free parameters, and thus greatly accelerate optimization problems. Differentiable models have been explored recently, for example, in non-coronagraphic modes of JWST \supercite{desdoigts2025amigo} and ground-based PSF reconstruction\supercite{Kuznetsov2026}. Preliminary experiments on simplified simulated coronagraphic data\supercite{feng2025exoplanet} have suggested that differentiable coronagraphic models could have a large impact in exoplanet imaging.}

\textcolor{black}{Here we present a framework for coronagraphic exoplanet imaging that leverages \textcolor{black}{differentiable models} for model-based PSF subtraction to deliver dramatic gains in exoplanet detection sensitivity. Our approach, Physics-based Learning for Imaging Planets (FLIP), incorporates a multi-step optimization strategy with a flexible physical model \textcolor{black}{informed by wavefront sensing data} to achieve the PSF fidelity necessary for real data. We demonstrate this framework with real JWST/NIRCam observations, providing the first successful application of differentiable model-based PSF subtraction on real on-sky coronagraphic data. Our approach improves exoplanet detection limits by up to $15\times$ relative to published analyses that used standard pipelines. These improvements enable the recovery of planetary signals that were previously inaccessible to existing approaches. We demonstrate this by obtaining the first image of giant planet 51 Eridani b at 4.3 $\mu$m, which previously resulted in a non-detection with standard techniques \supercite{balmer2025jwst}. These results show the potential of physics-based, autodiff-enabled algorithms to surpass limits set by telescope stability in current and future observatories. }

\subsection{FLIP method overview}\label{overview}

\begin{figure}[H]
    \centering
    \includegraphics[width=\linewidth]{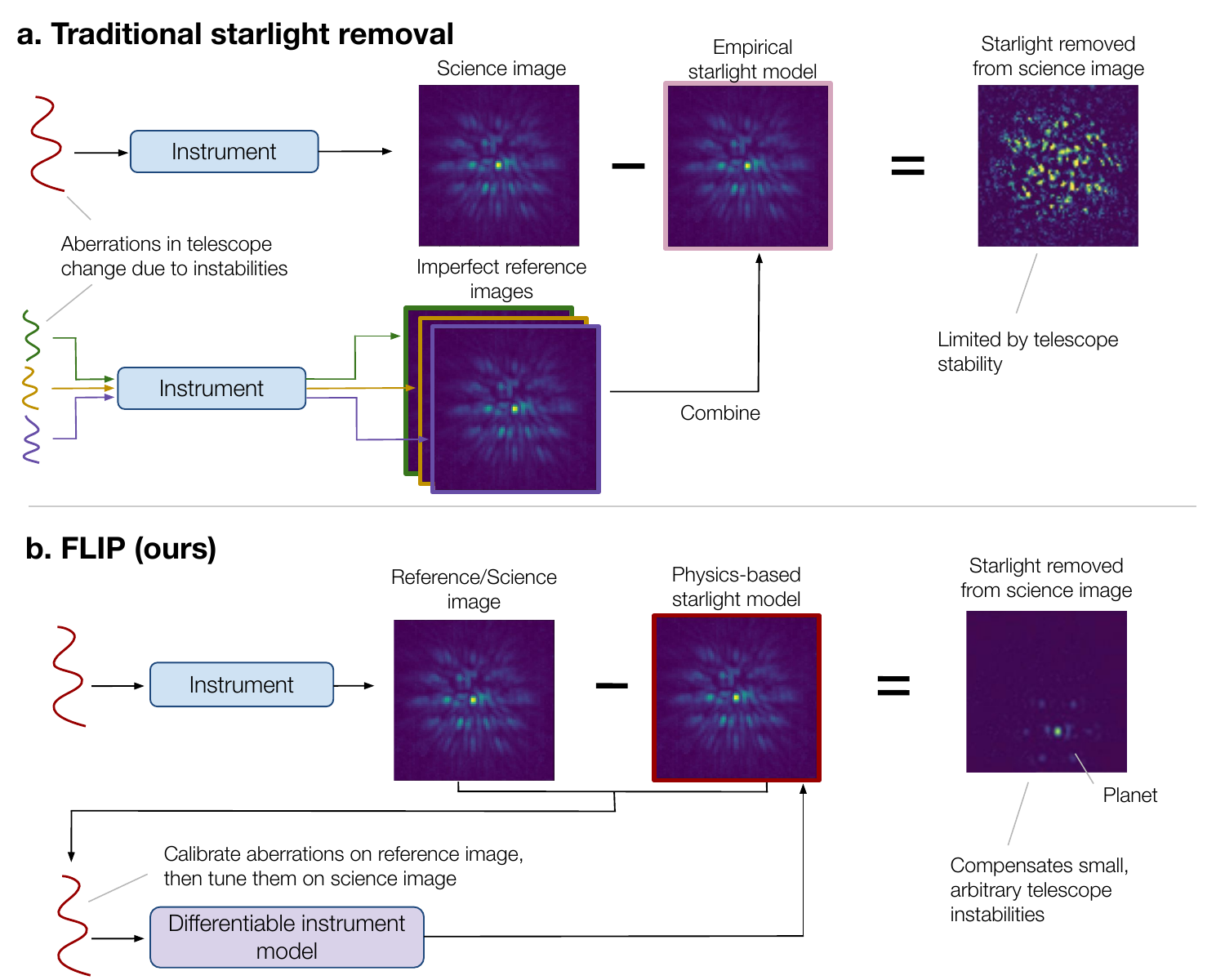}
    \caption{\textcolor{black}{\textbf{Schematic comparison of standard starlight subtraction \textcolor{black}{(top)} and our method based on physical modeling, FLIP \textcolor{black}{(bottom)}}. The stellar PSF is shaped by inevitable and imperfectly-known aberrations in the telescope (including star-coronagraph misalignment and optical phase aberrations, represented by colored curves to the left of the diagram). Starlight subtraction methods based on empirical reference images are limited by telescope instabilities that cause changes in these aberrations, since these result in dissimilar reference PSFs \textcolor{black}{(symbolized with curves and image outlines of different colors)} and thus imperfect PSF subtraction \textcolor{black}{when combining them into an empirical PSF model}. \textcolor{black}{Our method} FLIP pre-calibrates an instrument model on reference images to reproduce the reference PSF, and then solves for small changes in telescope aberrations that better explain the science image. \textcolor{black}{This is represented by the optimization loop that tunes aberration estimates with arbitrary flexibility (and thus closer to the effective aberrations that produce the science PSF, represented by the red curves), beyond the limited aberration diversity provided by imperfect reference PSFs in standard approaches}. This circumvents the stability limit of the instrument and enables imaging of fainter planets.}}
    \label{fig:optimization}
\end{figure}

FLIP (Figure \ref{fig:optimization}) \textcolor{black}{ applies three primary methodological innovations \supercite{feng2025exoplanet,desdoigts2025amigo,Kuznetsov2026} to on-sky space-based coronagraphic imaging for the first time: \textit{physics-based modeling}, \textit{automatic differentiation (autodiff)} and a \textit{multi-stage optimization framework}. \textcolor{black}{At a high level, FLIP} can be summarized as follows. We use a \textit{physics-based instrument model} \textcolor{black}{and wavefront sensing data} to generate a simulated PSF of the star. Using gradient descent, we tune the parameters in our instrument model to match the real PSF of the star as observed on real data, minimizing the difference between the simulated star PSF and the real JWST/NIRCam science image. \textit{Autodiff} makes optimization via gradient descent computationally efficient. In our \textit{multi-stage optimization} we pre-calibrate our instrument model on a reference JWST/NIRCam observation before tuning it on our science observation. Pre-calibration places our model close to an accurate solution, while tuning on the science image overcomes the differences between reference and science images due to instrument instabilities. To mitigate overfitting of real planet signals, we inject fake planets on the science image and track their signal-to-noise ratio (SNR) during our optimization, halting it when SNR reaches a maximum or plateaus.}

 Our \textcolor{black}{physical} model consists of \textcolor{black}{an astrophysical} scene (a point source corresponding to the host star), an optical model and a detector model. Our JWST optical models are based on the nominal models provided in the open-source software \texttt{STPSF} \supercite{perrin2012simulating}, the \textit{de facto} JWST PSF modeling tool maintained by the Space Telescope Science Institute.  These models are known to be imperfect, particularly for NIRCam coronagraphy, but they represent the current best characterization available of the optics and known aberrations. We implement this physical model in \texttt{PyTorch} \supercite{Paszke_2019}\textcolor{black}{, following \supercite{feng2025exoplanet}, to make it differentiable and leverage autodiff in our gradient descent optimization.} With this model as our initial condition, we solve for wavefront aberrations, the position and flux of the star and the pixel-level response of the detector. 

\textcolor{black}{Our multi-stage optimization framework is used to overcome model mismatch between our model and the real instrument, essentially improving the initial condition for our science image optimization. For pre-calibration we choose a reference star similar to our science observation, typically one of the empirical reference images that would be used in traditional PSF subtraction methods. During our optimization on the science image, our fake planets are injected close to the sensitivity limit or at similar separation and brightness as the known planet in the image if there is one. They also serve as injection-recovery tests to quantify signal loss and sensitivity corrections in the final PSF-subtracted image. In this stage, we also slow down our optimization (scaling down the loss, and thus the gradient, during our gradient descent)  and tune only a subset of the model parameters to further reduce the risk of planet overfitting. This end-to-end optimization, tuning around $10^6$ free model parameters \textcolor{black}{with an Adam optimizer\supercite{kingma2014adam}}, takes around 35 minutes in a regular NVIDIA A30 GPU. See Methods for details.}

\subsection{Results}\label{results}
We show the performance of FLIP on published JWST NIRCam data, and we compare it with the optimal PSF subtraction obtained in those publications with standard method KLIP \supercite{soummer2012detection}. We analyze observations of two star systems with known planets: HIP 65426 \supercite{chauvin2017discovery} first published in \supercite{carter2023}, and 51 Eridani \supercite{macintosh2015discovery} originally published in \supercite{balmer2025jwst}. We additionally process an observation of TYC 5899, a star from a planet search survey with no known companions \textcolor{black}{and with significant aberration drifts between reference and science observations} \supercite{bogat2025probing}. \textcolor{black}{We choose this target} to showcase the exoplanet discovery potential of FLIP \textcolor{black}{even when faced with long-term observatory instabilities. For HIP 65426 and TYC 5899 we use the observations in the F444W filter with the round 335R coronagraph mask, and for 51 Eridani we process two filters, F410M and F430M, both with the bar coronagraph (in the NARROW position)\supercite{Girard2022,kammerer2022performance}}. \textcolor{black}{This choice of experiments shows the performance of FLIP across a wide diversity of stellar spectral types, coronagraphs, observational filters, reference star observation strategies and levels of observatory stability.}  Figure \ref{fig:subtractions} shows comparisons of the PSF-subtracted observations using KLIP and our method FLIP.






 FLIP achieves superior performance in all observations relative to KLIP, with large improvements in most cases. We obtain around \textcolor{black}{1.6} times better SNR on HIP 65426 b, 2.1 times better SNR on 51 Eridani b in F410M, and a \textcolor{black}{SNR $=$ 6.3} detection of 51 Eridani b in F430M, which yielded a non-detection in the original publication (sensitivity was improved by a factor of $\sim$4 at the location of the planet). For TYC 5899, we inject fake planets \textcolor{black}{prior to PSF subtraction with both KLIP and FLIP} to illustrate the sensitivity improvement. One of the injections is shown in \textcolor{black}{Figure \ref{fig:subtractions}}, which we recover at high significance while it is far from being detected with KLIP. Additionally, there is a bright, nearby star in the TYC 5899 scene which is still visible after our PSF subtraction, but it has been distorted. See Methods for details.

Figure \ref{fig:contrastcurves} shows significant improvement in sensitivity to exoplanet detection compared to KLIP. \textcolor{black}{For example, in TYC 5899, FLIP produces $\sim15$ times better sensitivity than KLIP at 0.5 arcsec and $\sim5$ times at 1 arcsec, and in the F430M image of 51 Eridani, FLIP provides $\sim5$ times better sensitivity at 0.5 arcsec, and $\sim2.5$ times better at 1 arcsec.} \textcolor{black}{We also estimate a fiducial stability-limited sensitivity for KLIP, where we quantify the predicted KLIP sensitivity limit assuming realistic differences between reference and science observations due to the changing state of the observatory, and assuming no other noise sources. We simulate observations with \texttt{STPSF} on similar observing conditions as the data, including realistic star-coronagraph misalignments \supercite{Girard2022} and OPD drifts, and accounting for algorithm over-subtraction intrinsic to KLIP.  While simplified, we show that the KLIP sensitivity on real data is consistent with our estimated stability-limited KLIP performance \textcolor{black}{within a factor of 2}. In most cases, FLIP achieves performance well below this limit.}

\textcolor{black}{These results are illustrative of the potential of FLIP to aid exoplanet discovery and characterization with JWST. Our new \textcolor{black}{SNR $=$ 6.3} detection of 51 Eridani b in F430M is the first image of the planet at 4.3 $\mu$m, where the absorption band of carbon dioxide is located. This is a wavelength range inaccessible to ground-based measurements due to telluric contamination from the Earth's atmosphere. With CO$_2$ being a good tracer of atmospheric metallicity, future work will use this detection to measure the 4.3 $\mu$m photometry of the planet and carry out a new atmospheric characterization. This will help assess previous evidence for an elevated metallicity of about 5 times the stellar value \supercite{balmer2025jwst, rajan2017characterizing} and better understand the formation history of 51 Eridani b. Similarly, our images of TYC 5899 show how FLIP can enhance exoplanet discovery, overcoming even sensitivity limitations from using reference stars observed months apart from the science observation. Several current exoplanet discovery surveys are similar to the TYC 5899 dataset, with self-referenced observing strategies. These use other science targets as reference stars to increase observing efficiency, which can include data taken across many weeks and months, as opposed to dedicated, ideal contemporaneous reference stars. This observation of TYC 5899 in particular was taken four months after the reference stars observations. Even in this case, FLIP yields sensitivity gains of up to an order of magnitude. With more than 100 observations taken in similar observing programs (e.g., JWST programs GO 5835, SURVEY 6005), the exoplanet discovery potential of using FLIP in these surveys is large.}


\subsection{Future improvements}\label{limitations}

By being insensitive to instrument instabilities in principle, \textcolor{black}{model-based PSF subtraction could be capable of reaching sensitivities down to the fundamental noise limit} \supercite{stark2024paths}. FLIP, in general, achieves contrast levels much closer to the noise limit than standard method KLIP (Figure \ref{fig:contrastcurves}), but the photon noise is not reached in any observation yet. \textcolor{black}{Furthermore, while FLIP can effectively compensate for phase drifts and random star-coronagraph misalignments that limit KLIP, our models have not yet been systematically calibrated to retrieve accurate estimates of these instabilities with associated uncertainties.} Clear avenues for improvement include a more sophisticated detector model to process deep observations of very bright stars \supercite{goudfrooij2024algorithm, desdoigts2025amigo}, disentangling degeneracies between different model parameters, \textcolor{black}{a detailed exploration of algorithm hyper-parameters\supercite{redai2023giant}} reducing the computational cost of the method and a framework to handle large data volumes for ground-based applications. See Methods for details.

\subsection{Discussion}\label{discussion}


\textcolor{black}{After over thirty five years of development in coronagraphic exoplanet imaging, this work is the first time that model-based PSF subtraction surpasses standard PSF subtraction techniques, with up to an order of magnitude improvement in sensitivity.In the short term, the demonstrated $2-15\times$ sensitivity improvement will be significant for exoplanet discovery and characterization with JWST, enabling access to fainter, and thus colder, lower mass or older, giant planets in JWST/NIRCam observations \supercite{carter2023uncharted,biller2025imaging,carter2024into}.This will expand the imaged exoplanet population precisely in the region of parameter space where JWST is most valuable\supercite{carter2021direct}. Furthermore, with the predicted discovery of thousands of wide-separation giant planets from the Gaia mission\supercite{perryman2014astrometric, lammers2025exoplanet}, FLIP could be instrumental to characterizing some of these planets with JWST. In addition, our framework can also be extended to the JWST Mid-Infrared Instrument (MIRI), contributing to exoplanet detection and characterization in valuable wavelengths beyond 5 $\mu$m.}

\textcolor{black}{In the long term, these first results show the potential of physics-based image processing on other current and future telescopes, as well as the crucial role that machine-learning-based techniques can play. For ground-based telescopes, recent work \supercite{Kuznetsov2026} has shown great promise in non-coronagraphic PSF calibration. FLIP could be extended to coronagraphic ground-based imaging aided by detailed wavefront sensing and adaptive optics telemetry. While this involves distinct data volume challenges (see Supplementary Discussion), the potential to reach the photon noise limit could be extremely valuable (especially at shorter wavelengths where we are typically limited by contrast rather than thermal background), particularly to directly image temperate terrestrial planets around low mass stars with the extremely large telescopes \supercite{kasper2021pcs,males2024high}. Similarly for future space-based missions, the extreme levels of stability needed to image exo-Earths with the future HWO are a major challenge \supercite{nemati2024leveraging,coyle2024achieved}, 
aggressively restricting the design space for the telescope. In this work, we have shown sensitivity beyond the stability limit in JWST/NIRCam. If this can be extended to HWO contrast levels, the ultra-stability requirement for the system may be partially lifted, with the error budget necessary for image post-processing being reduced, and thus the design of the observatory may become more tractable. Investigating this through testing in the Roman Space Telescope \supercite{bailey2023nancy} and the Lazuli Space Telescope \supercite{roy2026lazuli} will be critical. If successful, the goal of imaging true exo-Earths and achieving photon noise-limited high contrast imaging for the first time will be closer to reality.}

\newpage

\begin{figure}[H]
    \centering
    \includegraphics[width=\linewidth]{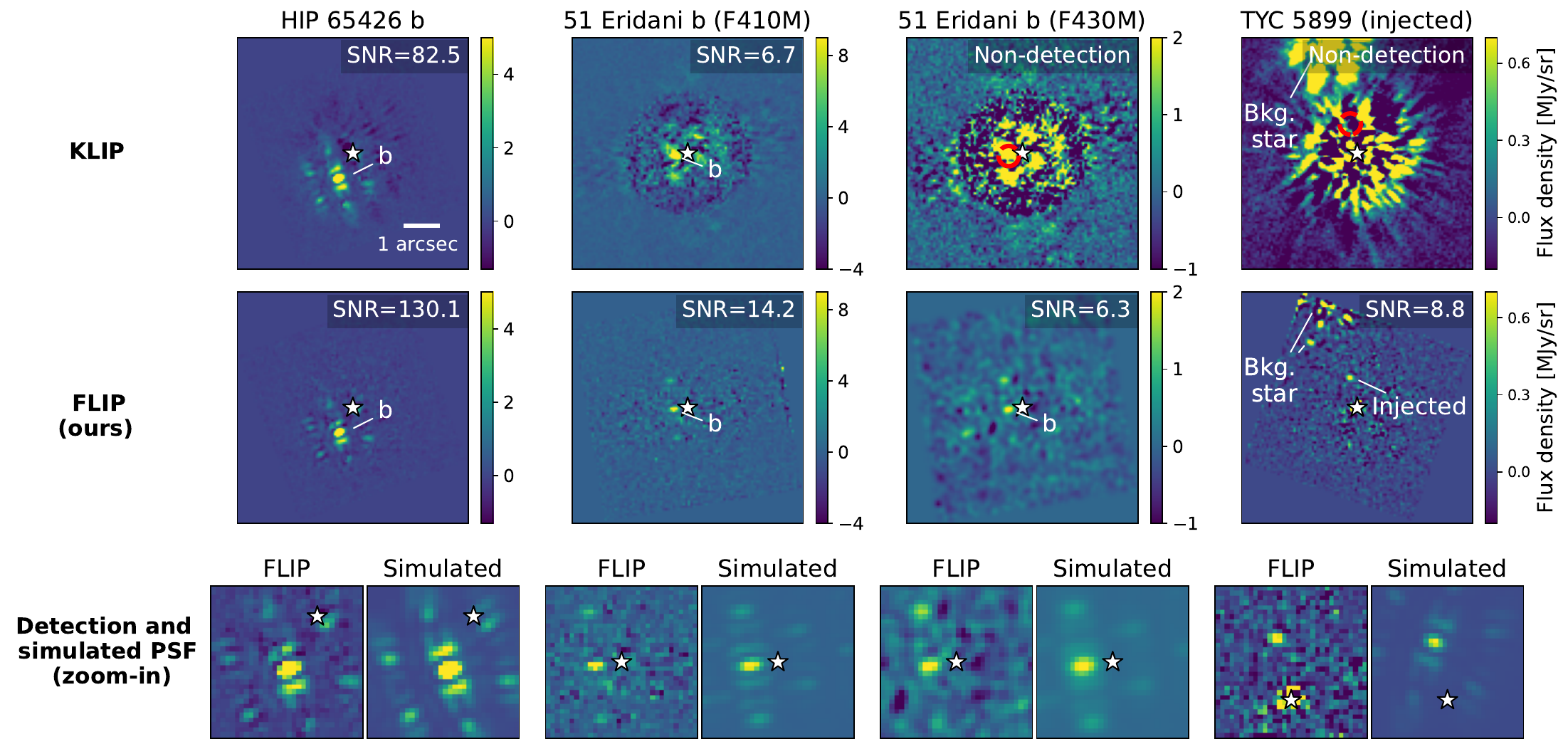}
    \caption{\textbf{Stellar PSF subtraction comparison between our method FLIP and standard method KLIP}. Top: \textcolor{black}{Observations re-processed in this work with KLIP, using the same algorithm hyper-parameters as in their respective publications. In all cases our re-processed KLIP results are comparable with the published results \supercite{carter2023, balmer2025jwst, bogat2025probing}}. Middle: Data processed with FLIP, showing the full field of view optimized over (80$\times$80 pixels). Bottom: Our detected companions with FLIP next to ideal simulated PSFs, to highlight the recovered morphology and to distinguish real point sources from PSF side lobes. The SNR of the detection is shown in the upper right corner of each panel. \textcolor{black}{Accounting for small sample statistics following \supercite{mawet2014fundamental}, the confidence of the detections from top left to bottom right correspond to 23.7$\sigma$, 5.7$\sigma$, non-detection, non-detection, 26.3$\sigma$, 8.8$\sigma$, 5.3$\sigma$ and 7.7$\sigma$}. The position of the star is marked with a star-shaped icon. In all panels, North is up and East is left. 51 Eridani F430M observations are convolved with a Gaussian kernel ($\sigma=$ 1 px) to smooth out pixel-to-pixel noise (only in this panel is pixel-to-pixel noise comparable with the brightness of the companion).}
    \label{fig:subtractions}
\end{figure}

\begin{figure}[H]
    \centering
    \includegraphics[width=\linewidth]{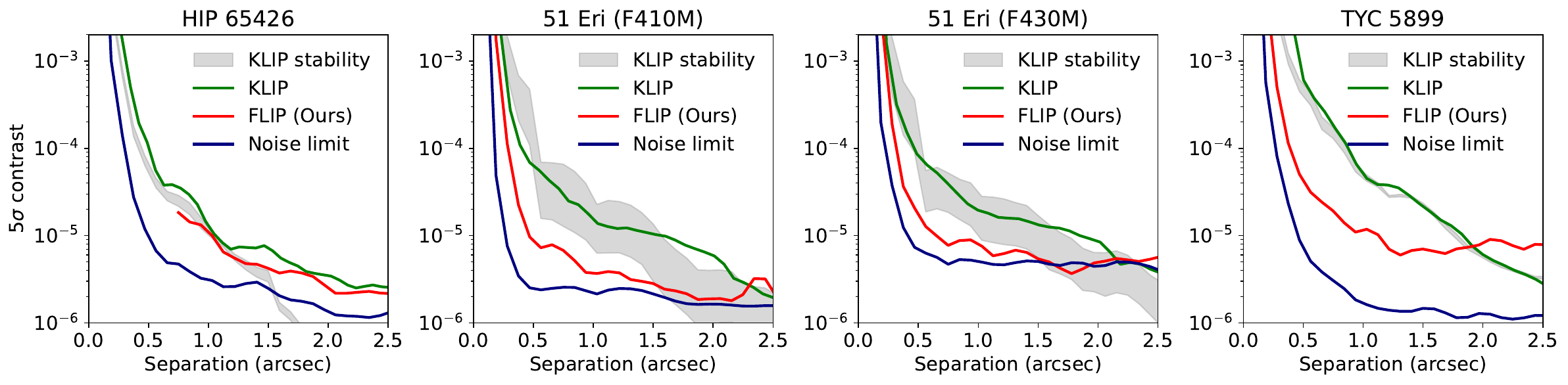}
    \caption{\textbf{Detection sensitivity curves for KLIP and FLIP on each on-sky dataset}. These curves are calibrated with injection-recovery tests. Also shown is the fundamental noise-limited sensitivity \textcolor{black}{for these on-sky datasets, and a range of estimated stability-limited KLIP sensitivity computed using simulated data.} Our method enables significant improvements (in some cases \textcolor{black}{$4-15\times$}) in exoplanet detection sensitivity. \textcolor{black}{For the stability limit, we simulate 50 random realizations of reference and science observations with realistic OPD drifts and star-coronagraph misalignment, and we run the same KLIP on these data (shaded regions correspond to the central 68\% confidence interval from these 50 random draws). } \textcolor{black}{For HIP 65426, we do not sample the FLIP sensitivity within 0.6 arcsec due to contamination from the bright planet PSF lobes, while the KLIP reductions do because they are computed after subtracting the planet with KLIP-FM \supercite{pueyo2016detection}.} }
    \label{fig:contrastcurves}
\end{figure}


\putbib[sn-bibliography]

\newpage

\section{Methods}\label{methods}

\subsection{The FLIP framework}\label{methods_framework}

\subsubsection{Model design}\label{methods_model}

FLIP models the stellar PSF by optimizing many of the free parameters in a physical model that influence the PSF, including the star being observed and the instrument used for the observation. We base our models on the nominal \texttt{STPSF} optical models, which are used to simulate PSFs in JWST for a variety of instruments and filters. Extended Data Figure \ref{fig:diagrams_params} shows a schematic diagram equivalent to the physical models used to simulate observations with the JWST NIRCam coronagraphs. 

Given these models, we find that four groups of free parameters are sufficient for PSF fidelity \textcolor{black}{in our experiments}:

\begin{itemize}
    \item The flux of the star and its precise position behind the focal plane mask of the coronagraph.
    \item The optical path difference, or OPD (a phase aberration) before the focal plane mask of the coronagraph.
    \item The OPD after the focal plane mask of the coronagraph.
    \item The response per pixel in the detector.
\end{itemize}

Physically, these parameters capture the imperfect alignment of the star with the coronagraph (telescope pointing errors), the imperfect alignment of the primary mirrors in JWST (deformations or tilts relative to the closest-in-time OPD measurement), the imperfectly-known OPD internal to specific instruments in JWST (in our case, the coronagraphs in the NIRCam instrument) and imperfectly-known detector calibrations. While other physical mechanisms are capable of affecting the PSF (such as pupil shear, manufacturing errors in the optics or more complicated detector physics), we find that our choice of free parameters is generally capable of effectively reproducing the effects of these other systematics and produce high quality PSF subtractions. A summary of our model parameters is shown in Extended Data Table \ref{tab:parametersummary}. 

\textbf{Implementation details.} Following \texttt{STPSF}, our model implements light propagation in the Fraunhofer approximation via Fourier optics. When modeling any specific observation, \textcolor{black}{we use \texttt{STPSF} to calculate the nominal instrument state and produce a simulated PSF, taking into account the instrument and filter, the closest-in-time OPD measurement to the reference star observation and the spectrum of the host star. We take this instrument state and we replicate it in \texttt{PyTorch} \supercite{Paszke_2019}}. See \supercite{perrin2012simulating} for details on the \texttt{STPSF} model construction, \textcolor{black}{and see our Supplementary Methods for details on our implementation.}

\subsubsection{Optimization design}\label{methods_optimization}

Previous work on simulated monochromatic data \supercite{feng2025exoplanet} found that given a near perfectly accurate physical model, it is possible to optimize it directly on science observations and produce excellent PSF subtraction without overfitting planet signals. However, in our experiments with real data we find that accidentally subtracting planet signals is indeed possible. This is likely due to \texttt{STPSF} models having substantial inaccuracies relative to real JWST instruments \textcolor{black}{and to our model being significantly more flexible with more free parameters}, thus making the optimization problem less straightforward.

To address this, we implement a two-stage optimization. In stage 1, we leverage the full flexibility of our model and we calibrate it on a reference star \textcolor{black}{(since, by design, it is unlikely to host any contaminants like planets or binaries)}, optimizing all the free parameters described in Section \ref{methods_model}. This helps overcome inaccuracies in the optical model and sets the parameters closer to a good solution. In stage 2, we use our calibrated model and optimize only a reduced number of free parameters directly on the science observation. This accounts for different pointing errors and different OPDs between reference and science observations due to telescope instabilities.

In both stages, we optimize different parameters in separate sub-stages. This is done to reduce degeneracies between different parameters. In stage 1 we have three sub-stages: 1) solving for star position and flux, 2) solving for both OPDs, and 3) solving for both OPDs and  pixel responses. In stage 2 we have two sub-stages: 1) solving for star position and flux, and 2) solving for the OPD after the focal plane mask. While the OPD before the coronagraph mask is also expected to change due to the JWST mirrors, we currently find the optimization to be more stable if we solve for the OPD after the mask instead \textcolor{black}{ (since this OPD does not need to be propagated through the coronagraph mask)}. These stages are summarized and illustrated in Extended Data Figure \ref{fig:optim_summary}\textcolor{black}{, with the number of iterations dedicated for each one. On a single NVIDIA A30 GPU, stage 1 takes 10 to 15 minutes, and stage 2 takes 10 to 30 minutes}.

\textcolor{black}{We carry out ``leave one out" tests to illustrate intuitively the reasons behind our contrast improvement (\textcolor{black}{Extended Data Figure} \ref{fig:leaveoneout1}). \textcolor{black}{In the top two rows of} \textcolor{black}{Extended Data Figure}  \ref{fig:leaveoneout1} we see that the star position has little effect in the contrast: while it is physically important (hence our inclusion), the effect is likely absorbed by other OPD terms in the model as tip-tilt phase aberrations. It is clear that OPDs strongly affect the PSF, since they are related to it directly via a Fourier transform. Detector responses are required to remove pixel-level errors that are not due to and cannot be fit by OPDs. This illustrates the source of the improvements relative to approaches like KLIP, since methods based on empirical reference images cannot account for arbitrary differences between reference and science data (stage 2.1 and 2.2). \textcolor{black}{The bottom row of} \textcolor{black}{Extended Data Figure}  \ref{fig:leaveoneout1} shows that model calibration is crucial, since removing it drastically increases the risk of planet over-subtraction. In \textcolor{black}{the bottom row of} \textcolor{black}{Extended Data Figure}  \ref{fig:leaveoneout1} we also see that a single OPD plane is not enough model flexibility: in our formulation, the interaction between both OPD planes (separated by a focal plane mask) is important. This illustrates the improvements of FLIP relative to prior autodiff-related efforts, notably \supercite{feng2025exoplanet} (which used a much less flexible model and no model calibration).}

\textbf{Implementation details.}  We use the Adam optimizer\textcolor{black}{ \supercite{kingma2014adam}} with a Cosine Annealing learning rate scheduler. We use an L1 loss function for our optimization. 

We select the learning rates in our model by hand, increasing it across orders of magnitude until we find one that produces a stable optimization with convergence below around 15000 iterations. Due to the different magnitudes of different parameters in our model, we assign a different learning rate to each family of parameters. A simple coarse search clearly shows the appropriate scale for each parameter. Extended Data Figure \ref{fig:lr_grid} shows order unity variations to learning rates do not significantly affect our results, and that increasing or decreasing them by orders of magnitude clearly shows the correct scale for stability and relatively fast convergence. A summary of the order of magnitude of these learning rates per parameter is shown in Extended Data Table \ref{tab:parametersummary}.

Across different observations, we find that globally scaling all learning rates by some order-unity constant can help with stability and convergence speed. This is likely due to the magnitude of the loss function depending on the brightness of the stars. We find that a global scaling of 4 works well in all the experiments presented here, but preliminary analysis on fainter stars showed better results by using 1 or 0.5 instead in those cases. 

Similarly, during the last sub-stage of our optimization, we reduce the global scaling factor to around 1$\%$ of its original value to further reduce risk of overfitting. Our results are not strongly sensitive to the exact choice of this value, but it can be tuned differently for different observations depending on the similarity between reference star and science target.

To determine a stopping point for our optimization, we inject fake planets in the science observation and we track their SNR as a function of iteration number. \textcolor{black}{When measuring sensitivity, we inject planets at 5-7 separations and 1-3 planets per separation. For each injection, we inject the fake planet by simulating an off-axis PSF with the nominal STPSF model at the desired location in the field of view and adding it to the science image. We inject them at $4-20$ times the 5$\sigma$ fundamental noise contrast limit. This number may be adjusted depending on the amount of signal loss observed. We inject each planet individually per FLIP run. We measure the SNR every $\sim$ 300 iterations. As expected, the SNR of each injected planet increases for a number of iterations and then decreases due to algorithm overfitting. Therefore we stop our optimization at the approximate iteration when the injected planets reach an SNR plateau or peak on average, which was found to be consistent for planets injected at similar separations. \textcolor{black}{Extended Data Figure}  \ref{fig:injections_snr_sl} shows this for fake planets injected at a variety of separations. In light of this, in our experiments for simplicity we choose our stopping point tracking a single fake planet injection. Just like with KLIP, the optimal fake planet injection strategy depends on context. For instance, one can choose as the stopping point the iteration number with maximum average SNR across all planet injections (e.g., if the goal is to maximize sensitivity across the field of view), or based on one or a few injections at a specific brightness and position (e.g., if the goal is to optimize the SNR of a specific planet signal). \textcolor{black}{Extended Data Figure}  \ref{fig:injections_snr_sl} also shows that our PSF subtraction is not strongly sensitive to our specific choices of fake planet injections. }

\subsection{Experiments}\label{methods_experiments}

We carried out experiments on four datasets: three with known companions (HIP 65426 in \textcolor{black}{the F444W filter} and 51 Eridani in \textcolor{black}{the F410M, F430M filters}) and one with no companions, part of a planet search survey (TYC 5899) \textcolor{black}{in the F444W filter}. \textcolor{black}{HIP 65426 and TYC 5899 are observed with the \texttt{MASK335R} coronagraph mask, while 51 Eridani is observed with the \texttt{LWBAR} coronagraph mask in the \texttt{NARROW} position offset. We selected this diverse set of data to be representative of a wide range of observations with JWST NIRCam coronagraphy. In particular, our four datasets include:}

\begin{enumerate}
    \item \textcolor{black}{Three different physical systems (HIP 65426, 51 Eridani, TYC 5899).}
    \item \textcolor{black}{Three different observational filters (F444W, F410M, F430M).} 
    \item \textcolor{black}{Two different coronagraphs (NIRCam 335R round coronagraph, and NIRCam LWBAR bar coronagraph).}
    \item \textcolor{black}{Both real planets (HIP 65416 b and 51 Eridani b) and simulated planets (TYC 5899).}
    \item \textcolor{black}{Observations with a dedicated reference star with a 9-POINT-CIRCLE dither pattern (HIP 65426), with a dedicated reference star with 5-POINT-BAR dither pattern (51 Eridani), and without dedicated reference stars but with a PSF library composed from a self-referenced survey (TYC 5899). This is relevant for the KLIP performance under a variety of reference star configurations.}
    \item \textcolor{black}{Observations with reference stars close in time to the science target (HIP 65426 and 51 Eridani) and far in time from the science target (TYC 5899).}
    \item \textcolor{black}{Observations of host stars of various spectral types: A2V (HIP 65426), F0IV (51 Eridani) and M3 (TYC 5899).}
\end{enumerate}

We run standard PSF subtraction method KLIP and our technique FLIP on all four datasets and compare the performance of each, both with the SNR of the known recovered companions and with the planet detection sensitivity at multiple separations from the star. We perform our KLIP subtractions in accordance with the optimal set of KLIP hyperparameters described in the respective publications \textcolor{black}{to replicate what the authors considered the best performance}. \textcolor{black}{In particular, all KLIP subtractions are done using RDI+ADI (see e.g., \supercite{carter2023}) on both roll angle observations of the science targets, which we then de-rotate and mean-combine. In all cases, we obtain PSF subtractions comparable to the published works. For FLIP, for simplicity, we perform it on only one roll per dataset. Future work could explore the differences in sensitivity between using FLIP only on one roll, FLIP on both rolls and mean-combining the de-rotated images, and FLIP on both rolls, using ADI on the resulting images, and mean-combining the FLIP+ADI results.}

For TYC 5899, the original publication\textcolor{black}{\supercite{bogat2025probing} presents two epochs: one with a single roll angle (the second failed), and one with two roll angles observed four months after the reference star sequences. We use the latter in this analysis for both KLIP and FLIP, to illustrate the gains in sensitivity when there is significant drift between reference and science observations. While \supercite{bogat2025probing} used all three observations for their contrast curves and we replicate it only with the latter two, we obtain comparable contrast performance: $\sim3\times10^{-6}$ at 2.5 arcsec and $\sim4\times10^{-5}$ at 1.0 arcsec for KLIP and FLIP, and $\sim3\times10^{-4}$ for KLIP and $\sim5\times10^{-4}$ for FLIP at 0.5 arcsec.}  \textcolor{black}{Since} the target \textcolor{black}{has} no known companions, we inject planets at various separations and brightnesses to compare the performance of KLIP and FLIP in a controlled setting. 

\textcolor{black}{\textcolor{black}{Extended Data Table}  \ref{tab:experiment} shows the optimization hyper-parameters used during our reductions with FLIP.}

\subsubsection{Data preparation}\label{methods_datareduction}

The observations of HIP 65426, 51 Eridani and TYC 5899 and their reference stars were processed with the JWST pipeline \textcolor{black}{(version 1.18.1)} and the community-developed tool \texttt{spaceKLIP} \textcolor{black}{(version 2.2.1)}. Data processing steps (which include ramp-fitting, converting from detector counts to physical flux units and bad pixel cleaning) and observation set up are detailed in the respective papers for those observations \supercite{carter2023, balmer2025jwst, bogat2025probing}. 

The JWST and \texttt{spaceKLIP} pipelines include bad pixel flagging and cleaning. However, imperfect bad pixel cleaning can cause more difficulties in our model-based PSF subtraction than in methods like KLIP. This is because hot or dead pixels, or variations in pixel responses, cannot be ``canceled-out" by subtracting observations with the same pixel variations. Therefore, bad pixels that persist even after the automated cleaning routines are identified by eye and masked during our optimization process. 


\subsubsection{Choice of reference stars}\label{methods_refstars}

Both HIP 65426 and 51 Eridani were observations carried out with dedicated reference stars: a bright star with likely no planets around it, of similar spectral type to the science target, observed right before or after the science observation. For these targets, we use as our calibration star (optimization stage 1, Extended Data Figure \ref{fig:optim_summary}) the same reference stars used for those observations: HIP 68245 for HIP 65426, and HD 30562 for 51 Eridani. \textcolor{black}{We use a single dither for model calibration with FLIP for simplicity, but all dithers are used to build the principal components for KLIP.} See \supercite{carter2023, balmer2025jwst} for details.

For TYC 5899, the observation was taken as part of a self-referenced program, where only science observations are taken, and a large number of these are put together to be used as a reference library for every other science observation. \textcolor{black}{For our KLIP subtractions, we use the full PSF library used in \supercite{bogat2025probing} to reproduce the result in their published work.} For an apples-to-apples comparison of FLIP and KLIP, we chose one of the observations in the self-referenced library as our calibration star \textcolor{black}{for the stage 1 of FLIP}. We select the HIP 17695 observation as it has the most similar spectral type to TYC 5899 from the targets in the library. See \supercite{bogat2025probing} for details.

\subsection{PSF subtraction performance metrics} \label{methods_performance}


The contrast curves for both KLIP and FLIP were calculated by dividing the image into small concentric annuli and computing the standard deviation within each annulus accounting for small sample statistics \supercite{mawet2014fundamental} and for planet signal loss due to both algorithm overfitting and coronagraph throughput. \textcolor{black}{For reference, the signal loss from algorithm overfitting near the location of the measured planets (in the same order as in \textcolor{black}{Figure}  \ref{fig:subtractions}) for KLIP is $\sim5-10\%$,$\sim20-70\%$,$\sim50-80\%$, and $\sim10-40\%$ and for FLIP it is $\sim5-20\%$,$\sim20-40\%$, $\sim30-50\%$ and $\sim40-60\%$.} We also smooth all datasets with a Gaussian kernel ($\sigma=1$ px) before computing the standard deviations, since FLIP residuals are mostly dominated by pixel-to-pixel noise. For HIP6546, the FLIP contrast curve excludes the central 0.6 arcseconds of the image given that a bright lobe of the planet PSF prevents sampling this region, while KLIP uses the KLIP-FM algorithm to subtract the planet before computing sensitivity. Future work incorporating planet modeling and subtraction with FLIP will overcome this limitation. See \supercite{carter2023, balmer2025jwst, bogat2025probing} for details on computing and calibrating these contrast curves.

For the fundamental sensitivity curve, we use the \texttt{ERR} extensions of the JWST data products, which track the combined total uncertainty in each pixel value from read noise\textcolor{black}{, Poisson noise, and flat noise}. We generate a random Gaussian sample with mean 0 and $\sigma$ equal to the total noise level in each pixel. \textcolor{black}{That is, when generating our noise realization we are simulating what the image would look like after perfect PSF subtraction. } We finally compute the contrast curve on this image in the same way as described above for KLIP and FLIP subtractions, assuming no planet signal loss. \textcolor{black}{This assumes that the only source of fundamental noise comes from the science observation. This is essentially the case 0 presented in \supercite{godoy2024escape}. In KLIP, additional noise comes from the reference observations too (see \supercite{godoy2024escape}). FLIP is model-based PSF subtraction, so our FLIP subtraction adds no additional fundamental noise. The fundamental noise curves can also be interpreted as a lower limit, or the most optimistic case, of the possible sensitivity. Since we perform FLIP on only one roll per dataset, the noise limit corresponds to the noise limit of that roll.}

\textcolor{black}{To supplement Figure \ref{fig:subtractions} we compute SNR maps for each PSF subtracted image (\textcolor{black}{Extended Data Figure}  \ref{fig:SNRmapsFig2}). We generate a non-coronagraphic PSF model cropped to the central $5\times5$ pixels and we cross-correlate our image with the PSF model. Then we compute the signal-to-noise of each individual pixel within an annulus of width $\sim0.2$ arcsec in a two step process: we compute the noise per annulus to identify pixels with SNR$>5$ as a first pass, and then compute a refined noise estimate excluding those pixels. The final SNR map is simply the pixel values of the cross-correlation map divided by the refined noise estimate.}


\subsection{Estimation of the stability-limited sensitivity}\label{methods_stability}

\textcolor{black}{In Figure \ref{fig:contrastcurves} we show a fiducial stability-limited sensitivity. This limit is computed by performing KLIP on simulated data, where we can isolate the physical factors that contribute to the KLIP performance limitations. We perform 50 realizations of each observation sequence, stochastically varying the star-coronagraph misalignment and the OPD drift consistent with empirical JWST characterization in flight \supercite{Girard2022,telfer2024empirical}, as detailed below. To isolate the effects of these instabilities, we include no other source of noise (e.g., photon noise, flat noise) in our simulations.}

\textcolor{black}{For the science target, we generate \texttt{STPSF} models using the spectral type of the star, and we scale the flux within the central 20 pixel-radius to the flux of the real data. We simulate two rolls offset by 10 degrees in position angle. For the reference stars, for HIP 65426 we simulate 9 PSFs following the 9-POINT-CIRCLE dither strategy, for 51 Eridani 5 PSFs following the 5-POINT-BAR dither strategy} (\url{https://jwst-docs.stsci.edu/jwst-near-infrared-camera/nircam-operations/nircam-dithers-and-mosaics/nircam-subpixel-dithers/nircam-small-grid-dithers#gsc.tab=0}) \textcolor{black}{, and for TYC 5899, we simulate the same number of reference stars that were used in the PSF library of \supercite{bogat2025probing}. }

\textcolor{black}{We assume that all reference stars are of identical spectral type and brightness to the science target, and we assume no photon, detector or flat noise. In some targets, this results in the KLIP stability limit being deeper than the fundamental noise limit at large separations from the star (Figure \ref{fig:contrastcurves}). This is acceptable for our purposes, since at these close separations we are generally limited by PSF subtraction inaccuracies rather than fundamental noise, and we are mostly interested in improving sensitivity in the regions of the image where we are stability-limited rather than fundamental noise-limited.}

\textcolor{black}{For each observation, we simulate a target acquisition error with a random star-coronagraph misalignment drawn from $\mathcal{N}(\mu=0,\sigma=10)$ milliarcseconds, consistent with target acquisition accuracy in flight \supercite{Girard2022}. We simulate the PSFs with the closest-in-time OPD measurement. We assume that there is no significant OPD drift between reference star dithers. To mimic small OPD drifts due to telescope slews, for each observation in the real data that required a telescope slew (e.g., going from reference star to the science target) we add an OPD drift term to the OPD measurement. The shape of the OPD drift is taken from subtracting the two closest-in-time OPD measurements to the observation being simulated. The root-mean-square RMS (i.e., the amplitude) of the drift term is randomly drawn from $\mathcal{N}(\mu=4,\sigma=1)$ nm, smaller than drifts measured over day timescales ($\sim7-9$ nm RMS) and bigger than drifts on minutes timescales ($\sim1-2$ nm RMS) \supercite{telfer2024empirical}.}

\textcolor{black}{For HIP 65426, the bright companion has a non-negligible effect on PSF subtraction by skewing the KLIP algorithm (e.g., during mean subtraction and in the calculation of KL modes). We recreate this effect by injecting a fake planet in our HIP 65426 observation at the same contrast, separation and position angle as the real companion in our simulated observation.}

\textcolor{black}{We perform KLIP with ADI+RDI on all targets, using all KL modes for HIP 65426 and 51 Eri, and 6 KL modes for TYC 5899, following the respective publications. We compute our calibrated stability-limited KLIP sensitivity with injection-recovery tests, injecting planets at 3-7 separations and 2-3 position angles across the $80\times80$ px field of view, with the injected flux being 4 times the uncalibrated KLIP sensitivity at the injection separation. The central 68\% range of the resulting contrast curves are plotted as shaded regions in Figure \ref{fig:contrastcurves}.}

\subsection{TYC 5899 synthetic injections}\label{methods_fakeplanets}

To compare the sensitivity of FLIP and KLIP on TYC 5899, we inject \textcolor{black}{fake} planets with known flux across the image and assess the recovery PSF subtraction. Extended Data Figures \ref{fig:ourgallery} and \ref{fig:klipgallery} show a range of planet injections, complementing Figure \ref{fig:subtractions}. At each separation we inject a planet with flux equal to \textcolor{black}{20} times the \textcolor{black}{$5\sigma$} fundamental noise limit. 

For PSF subtraction, we process all observations in Extended Data Figures \ref{fig:ourgallery} and \ref{fig:klipgallery} identically, using the same learning rates and numbers of iterations for FLIP, and the same numbers of annuli and KL modes for KLIP. \textcolor{black}{This is a standard method to quantify sensitivity in KLIP (e.g., \supercite{redai2023giant}): once a set of annuli and KL modes have been chosen (e.g., to recover the signal in Figure \ref{fig:subtractions}), we inject fake planets across the field of view while keeping the annuli and KL modes fixed. This estimates the detection sensitivity across the field of view, given a set of hyperparameters. In FLIP, we adopt an analogous strategy, by injecting fake planets across the field of view, while keeping the learning rates and numbers of iterations fixed.}

We find that FLIP outperforms standard method KLIP in the vast majority of cases, especially at closer separations, producing higher SNR on a diverse range of possible planets. \textcolor{black}{We also see that in this particular dataset, at larger separations FLIP produces higher signal loss than KLIP in some cases, which explains the contrast curve for KLIP being deeper than FLIP at larger separations in Figure \ref{fig:contrastcurves}. Future work could address these differences in signal loss. Nevertheless, at these larger separations we are often times background-limited rather than contrast-limited; that is, sensitivity is mainly limited by exposure time rather than by starlight subtraction. In the interesting regime where we are limited by telescope stability and imperfect starlight subtraction, at smaller angular separations (which also correspond to more Solar System-like physical separations), FLIP robustly outperforms KLIP.}

\subsection{Data availability}\label{data}
The JWST NIRCam data used in this paper are from JWST programs ERS 1386 (PI: S. Hinkley), GTO 1412 (PI: M. Perrin) and GTO 1184 (PI: J. Schlieder) and are publicly available from the Mikulski Archive for Space Telescopes (MAST).

\subsection{Code availability}\label{code}
\textcolor{black}{The FLIP code and an example demo with test data are available at \url{https://github.com/rferrerc/pyflip/tree/submission} (submission commit: a92e684).}

\subsection{Acknowledgements}\label{acknowledgements}
\textcolor{black}{We thank W. Balmer, E. Bogat and A. Carter for assistance on processing their observations. We thank B. Pope, L. Desdoigts and M. Ygouf for useful discussions. This work is based on observations made with the NASA/ESA/CSA James Webb Space Telescope. The data were obtained from the Mikulski Archive for Space Telescopes at the Space Telescope Science Institute, which is operated by the Association of Universities for Research in Astronomy, Inc., under NASA contract NAS 5-03127 for JWST. These observations are associated with program 7942 (PI: Ferrer-Chavez). Support for programs 7942 was provided by NASA through a grant from the Space Telescope Science Institute, which is operated by the Association of Universities for Research in Astronomy, Inc., under NASA contract NAS 5-03127. This research was supported in part through the computational resources and staff contributions provided for the Quest high performance computing facility at Northwestern University which is jointly supported by the Office of the Provost, the Office for Research, and Northwestern University Information Technology. This work used computing resources provided by Northwestern University and the Center for Interdisciplinary Exploration and Research in Astrophysics (CIERA). This research made use of POPPY, an open-source optical propagation Python package originally developed for the James Webb Space Telescope project \supercite{perrin2012simulating}. }

\subsection{Author contributions}\label{contributions}
\textcolor{black}{R. F.-C., B. Y. F., K. L. B. and J. J. W. conceived the method presented in this work. R. F.-C. led the analysis, the writing of the manuscript and the writing of the FLIP code. J. J. W., K. L. B. and W. T. F. provided guidance on the method development and analysis. B. Y. F. contributed the initial version of the code demonstrated originally on simulated monochromatic data. L. P. and M. D. P. provided expertise on the JWST spacecraft, the NIRCam coronagraphs and STPSF models. All authors contributed feedback on the manuscript.}

\subsection{Funding statement}\label{funding}
\textcolor{black}{R. F.-C. and J.J.W. disclose support for the research of this work from the Space Telescope Science Institute [JWST-AR-07942]. J.J.W. discloses support for the research of this work from the Sloan Foundation [FG-2025-24976]. K.L.B. acknowledges support from a Sloan Fellowship and an Amazon AI4Science Discovery Award. W.T.F acknowledges support from the National Science Foundation under Cooperative Agreement PHY-2525568 (The NSF AI Institute for Artificial Intelligence and Fundamental Interactions, http://iaifi.org/).}

\subsection{Competing interests}\label{compiting_interests}
\textcolor{black}{Authors declare no competing interests.}

\subsection{Contact details for
corresponding author(s)}\label{Contact}
\textcolor{black}{Email: rodrigoferrer-chavez2029@u.northwestern.edu}


\subsection{Additional information}

\textcolor{black}{Supplementary Information is available for this paper.}

\textcolor{black}{Correspondence and requests for materials should be addressed to Rodrigo Ferrer-Chavez.}

\section{Extended Data}\label{extendeddata}

\setcounter{figure}{0}

\captionsetup[figure]{name=Extended Data Figure}

\begin{figure}[h!]
    \centering
    \includegraphics[width=\linewidth]{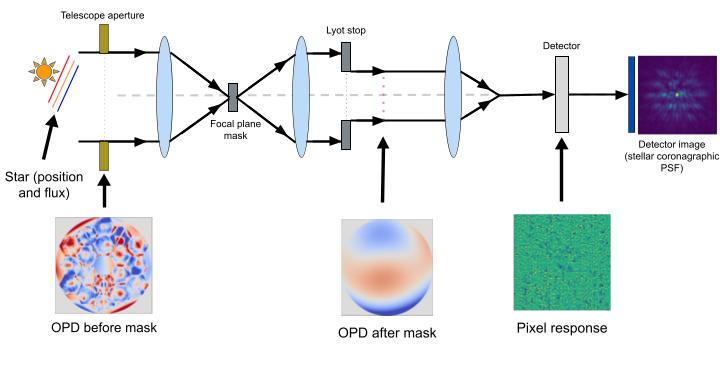}
    \caption{\textcolor{black}{\textbf{Schematic of the JWST coronagraph optical model used in this work}. We highlight the parameters we tune in our model.} }
    \label{fig:diagrams_params}
\end{figure}

\begin{figure}[h!]
    \centering
    \includegraphics[width=\linewidth]{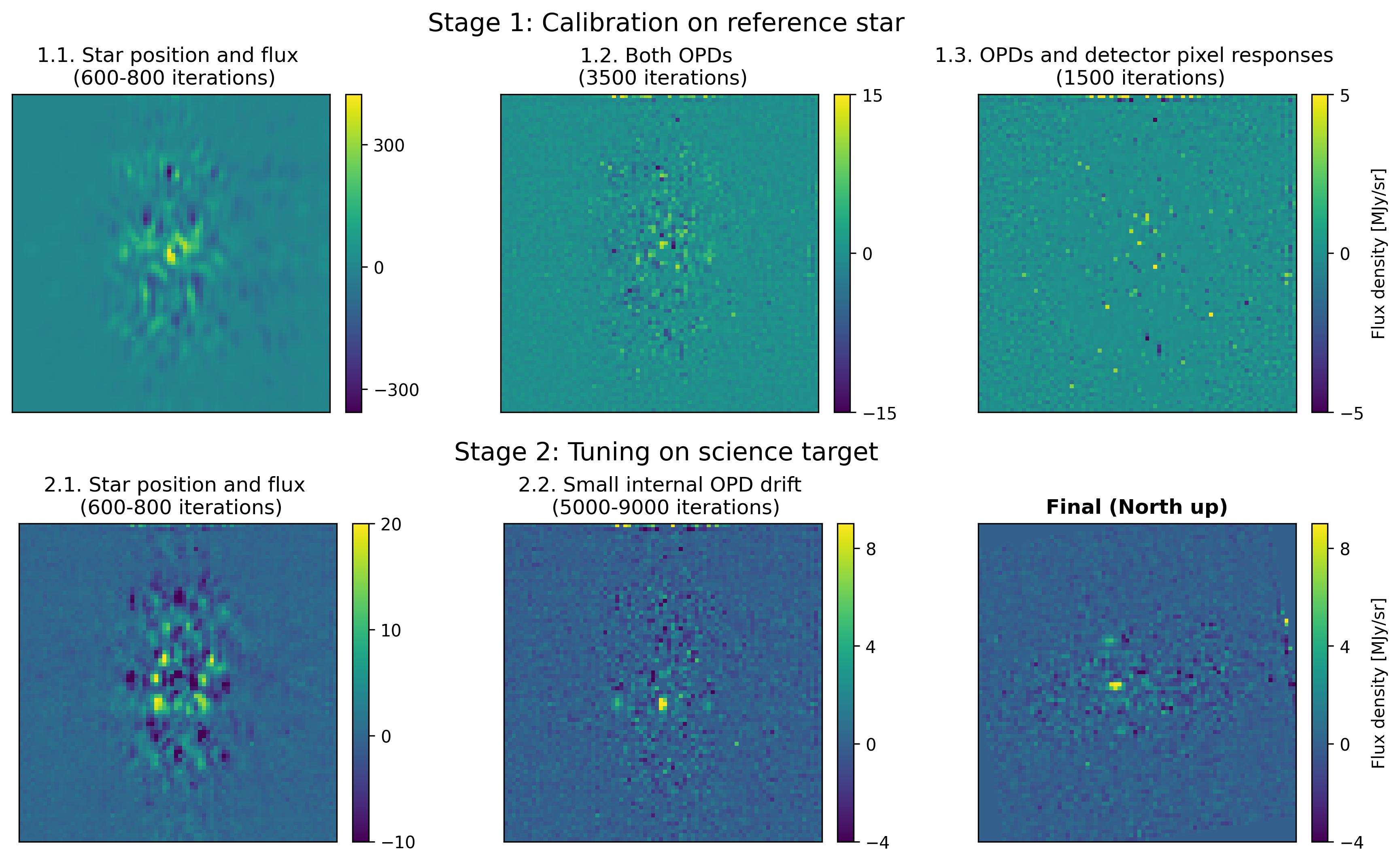}
    \caption{\textcolor{black}{\textbf{Summary of optimization stages in FLIP}. This includes calibration (stage 1\textcolor{black}{, top panels}) on a reference star and parameter tuning (stage 2\textcolor{black}{, bottom panels}) on the science target of interest. The color scale is adjusted in every step to show the dominant features of the residuals. The example shown is 51 Eridani in the F410M filter, revealing 51 Eridani b in the last panel. For visualization purposes we keep all panels in the detector orientation, except for the last one where we rotate it North up as in Figure \ref{fig:subtractions}. On a single NVIDIA A30 GPU, stage 1 takes 10 to 15 minutes, and stage 2 10 to 30 minutes.}}
    \label{fig:optim_summary}
\end{figure}

\begin{figure}
    \centering
    \includegraphics[width=\linewidth]{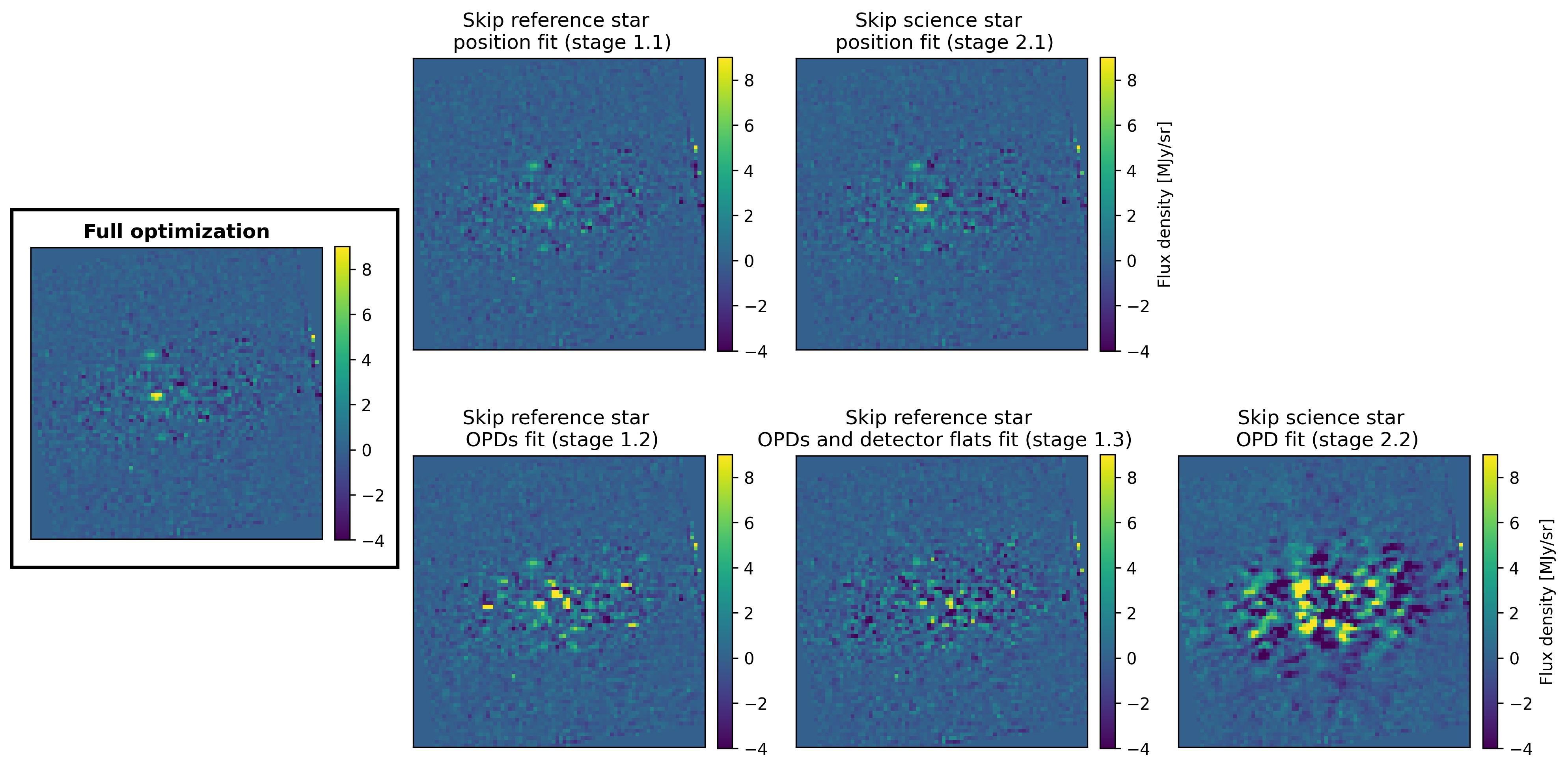}
    \includegraphics[width=\linewidth]{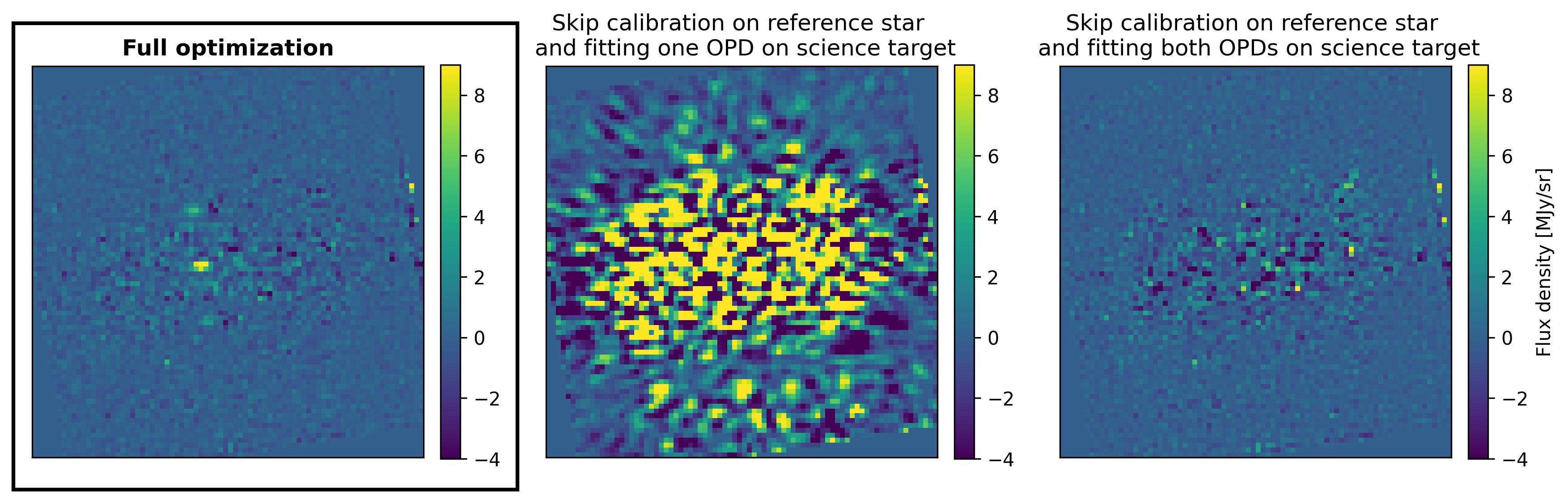}
    \caption{\textcolor{black}{\textbf{``Leave one out" tests on different stages of our optimization framework.} \textcolor{black}{Our full optimization is shown in the far left on 51 Eri (F410M) for comparison}. Top two rows: Excluding individual stages. We skip the fit for star-mask misalignment (top) and the fits for individual OPDs and for detector effects (bottom). In all cases we fit the flux of the star. Bottom row: Excluding the calibration stage on the reference star as a whole. We show the result fitting only for the post-coronagraph OPD (middle) and for both OPDs (right) on the science target. }}
    \label{fig:leaveoneout1}
\end{figure}


\begin{figure}[h!]
    \centering
    \includegraphics[width=\linewidth]{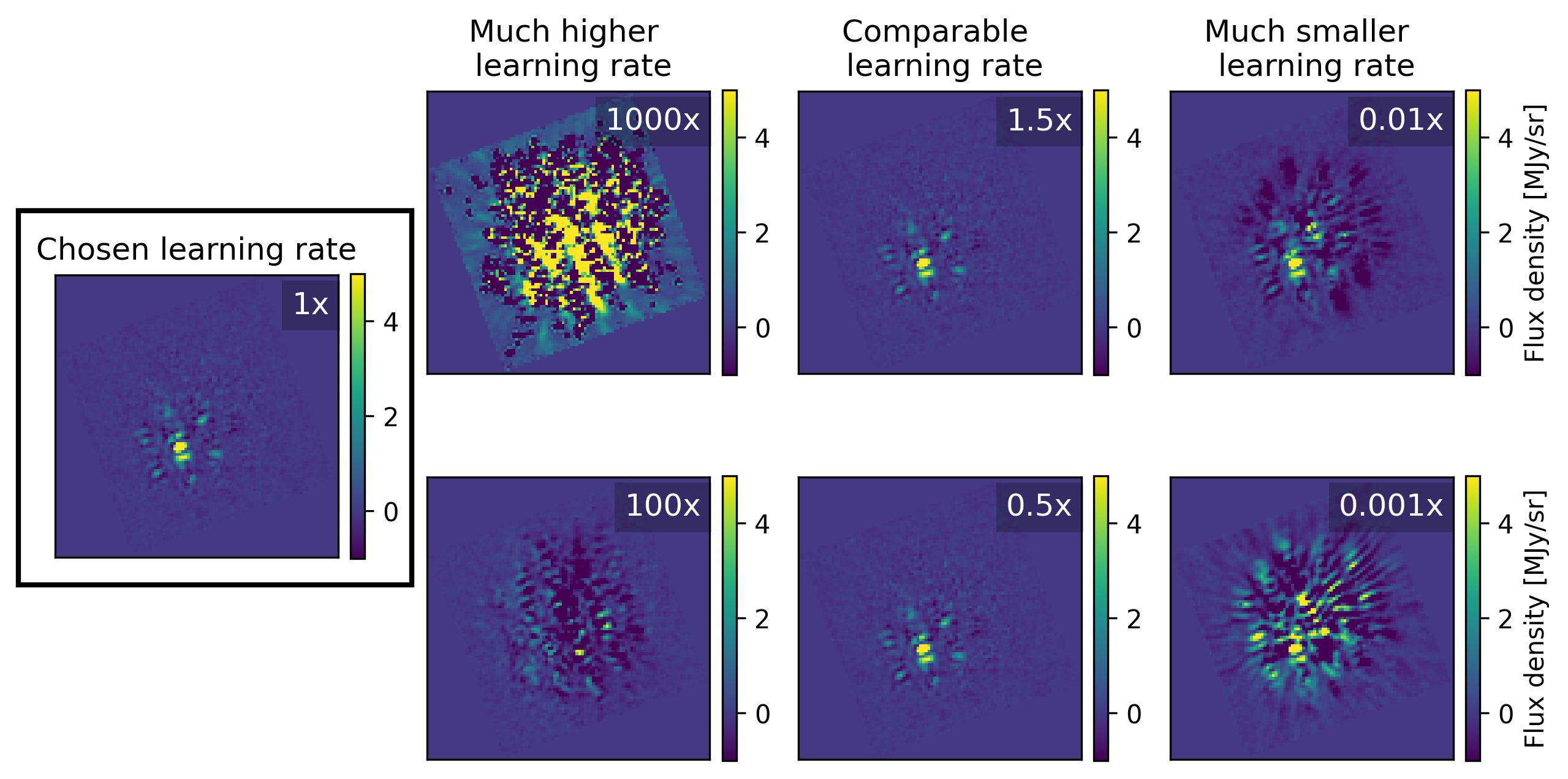}
    \caption{\textbf{Sensitivity to chosen learning rates}. Example FLIP runs using different learning rates while keeping number of iterations identical. \textcolor{black}{Increasing our learning rate by orders of magnitude leads to unstable optimization unable to converge (top left) or to drastic overfitting (bottom left). Lower our learning rate by orders of magnitude increases the amount of un-subtracted starlight (bottom right) or, for this specific target, reduces the ability of our model to compensate for imperfect stellar flux estimate (stage 2.1) due to the presence of the bright companion (top right).} In each panel, all learning rates in the model are scaled by the factor indicated inside the panel.}
    \label{fig:lr_grid}
\end{figure}

\begin{figure}
    \centering
    \includegraphics[width=0.48\linewidth]{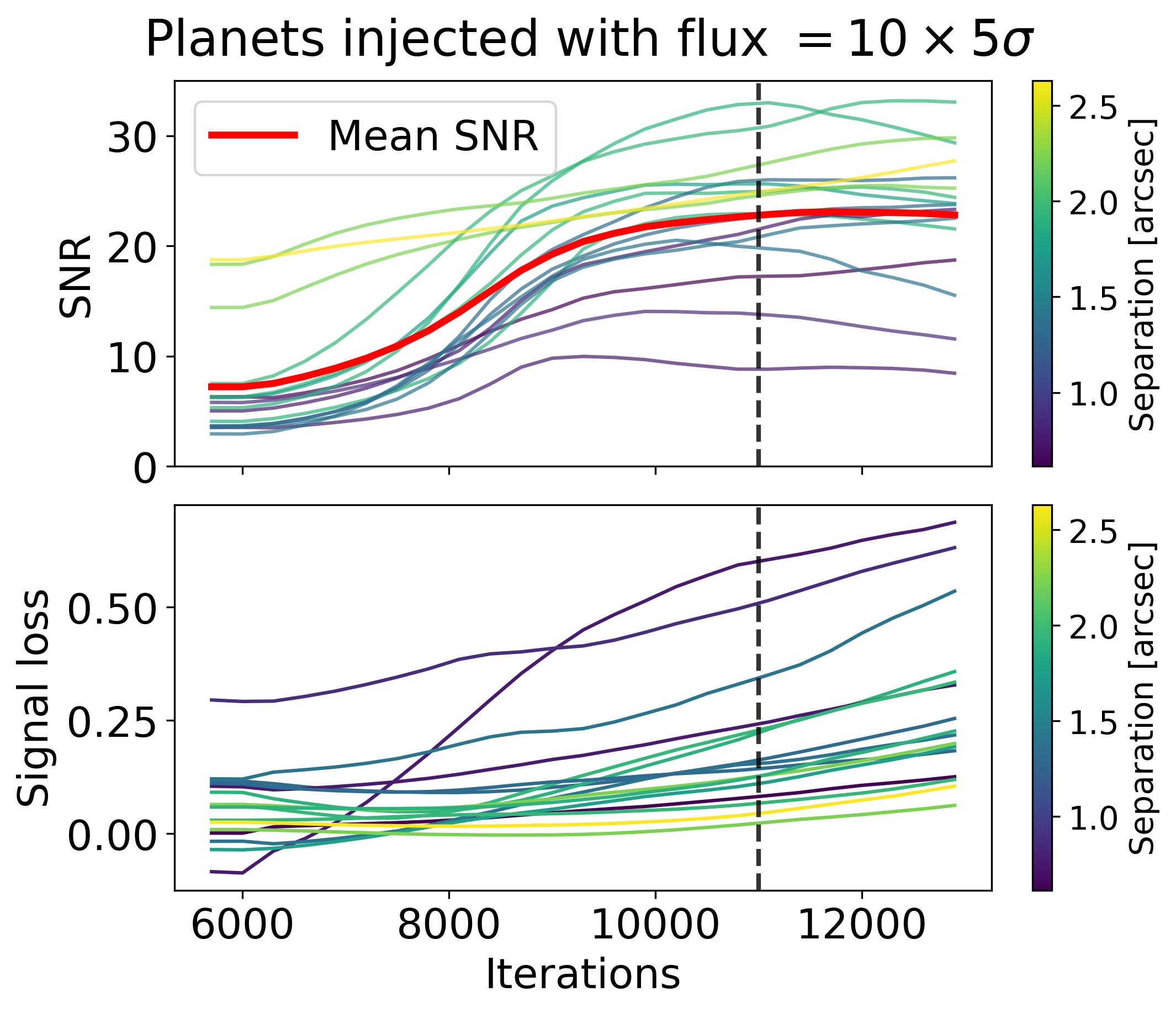}
    \includegraphics[width=0.48\linewidth]{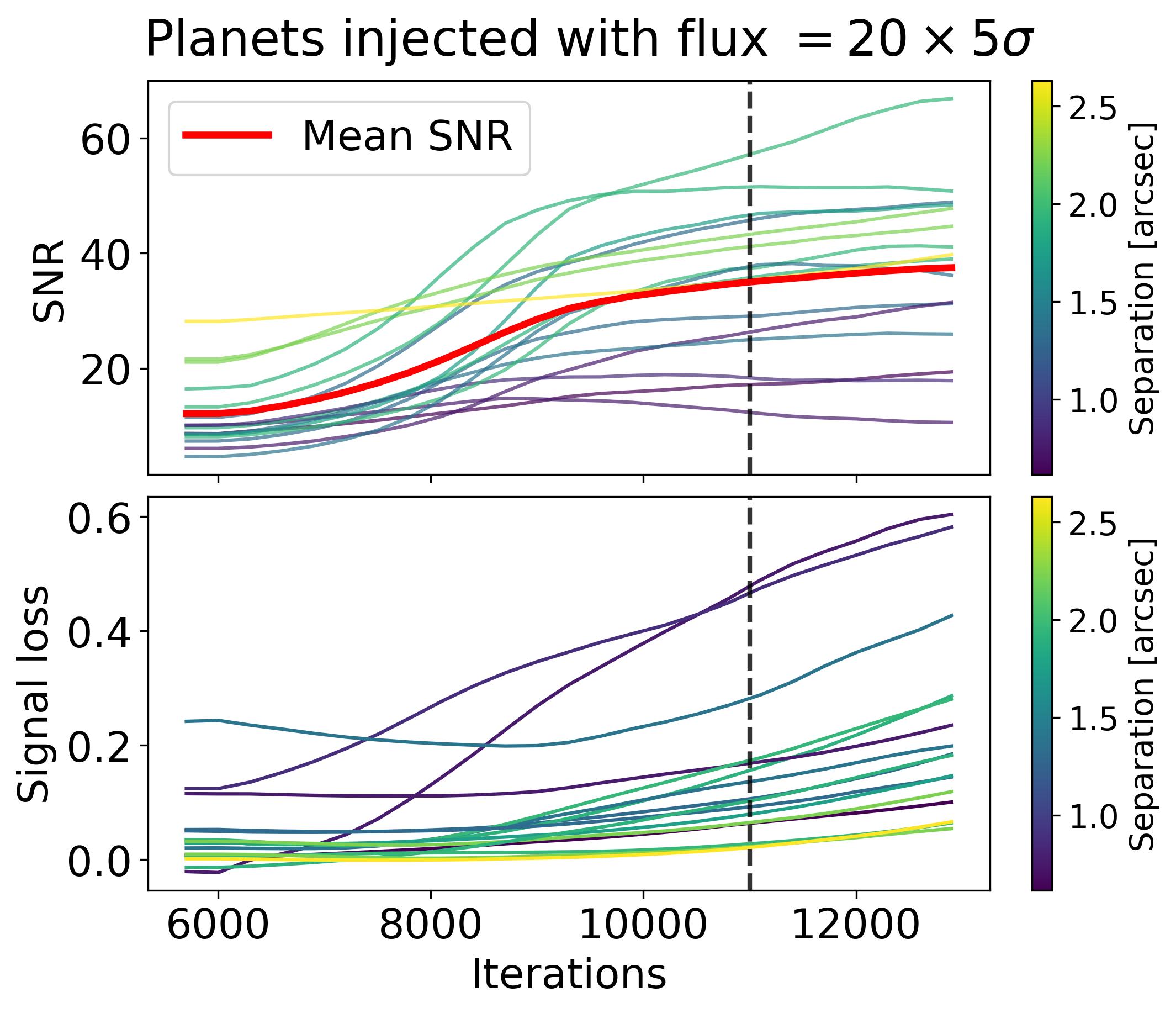}
    \includegraphics[width=0.48\linewidth]{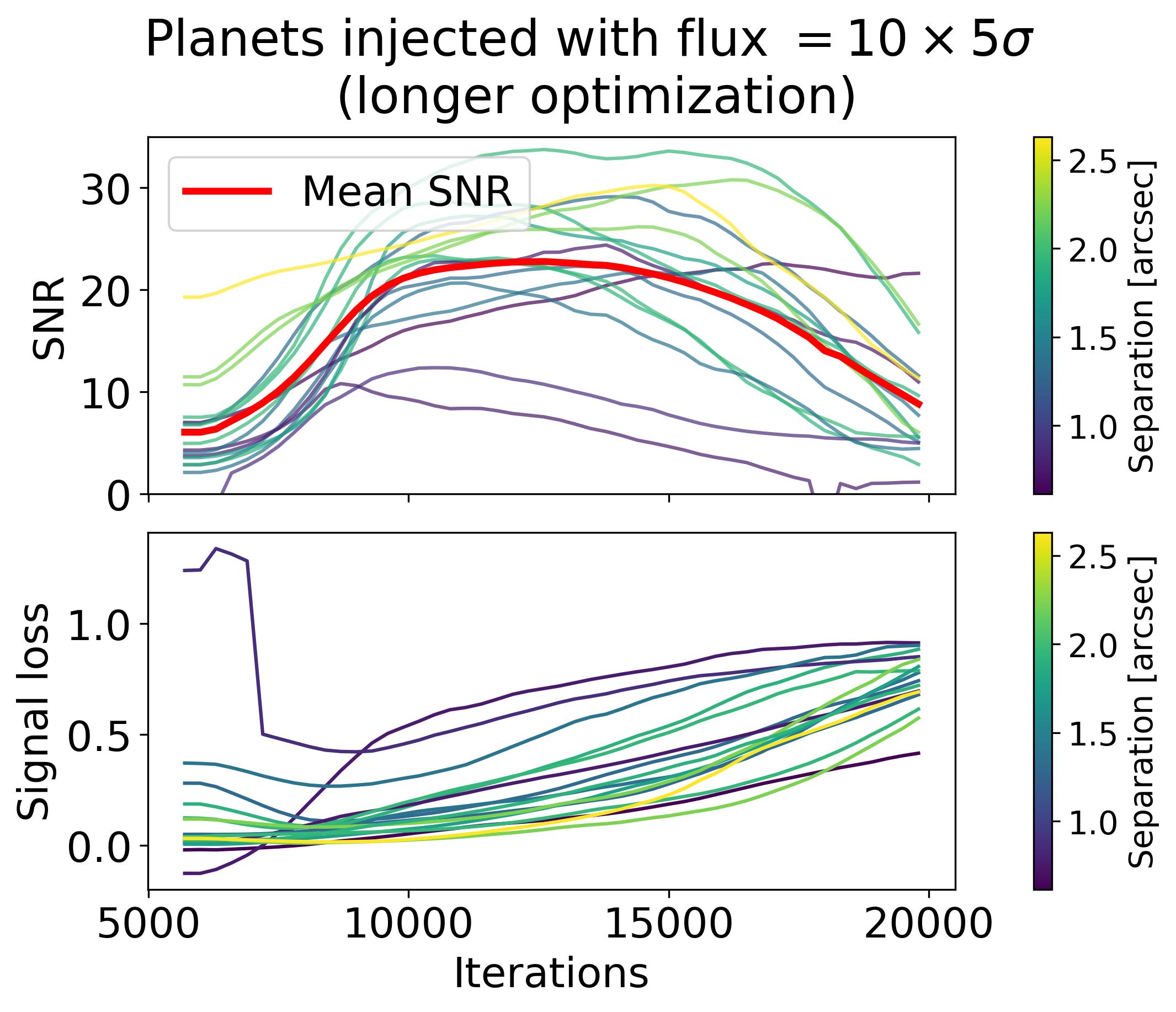}
    \includegraphics[width=0.48\linewidth]{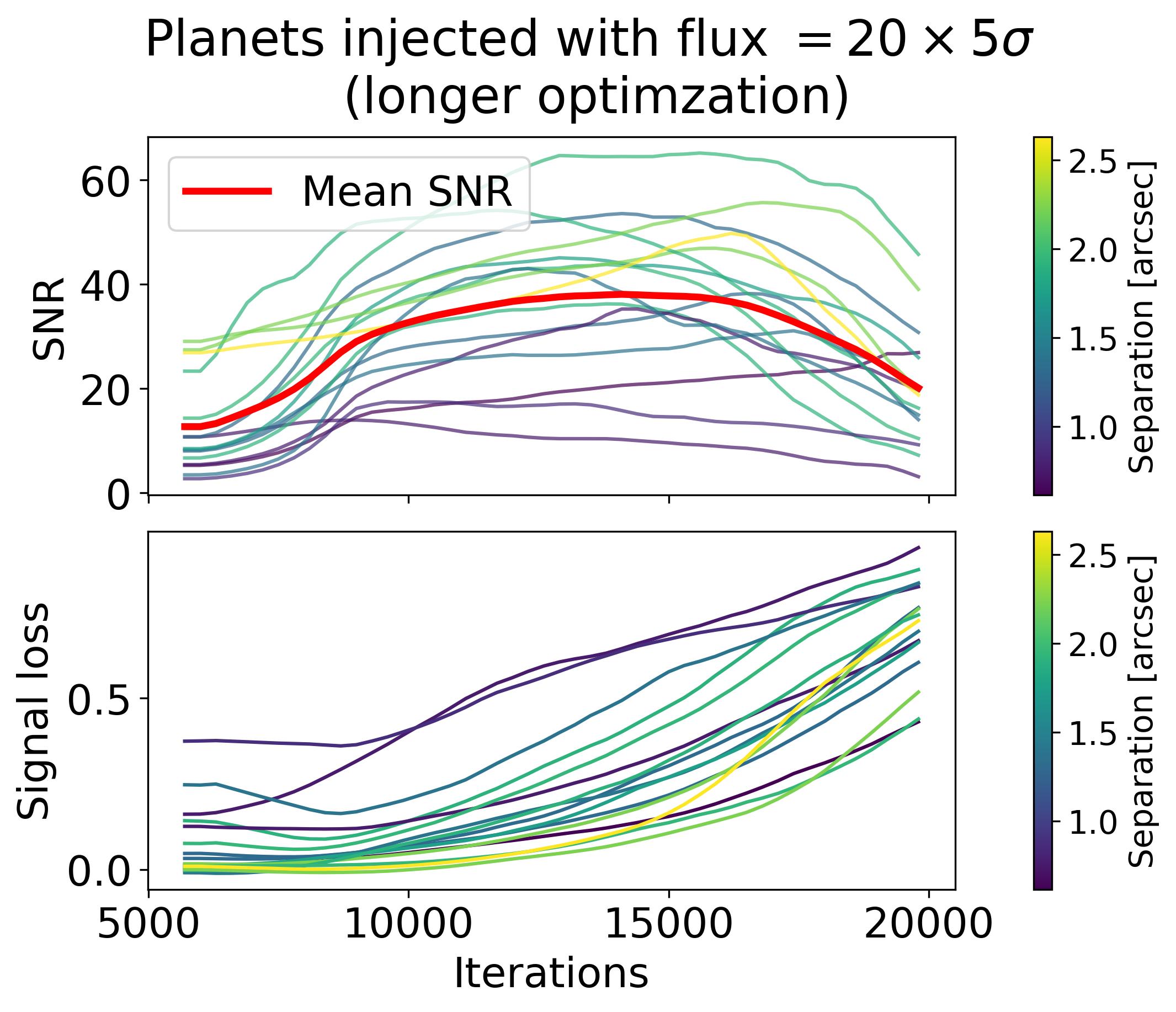}
    \caption{\textcolor{black}{\textbf{Signal-to-noise ratio and signal loss for 16 example fake planets injected in the HIP 65426 dataset}. In the left and right panels, all planets were injected at 10 and 20 times the 5$\sigma$ fundamental noise contrast curve (blue curve in the HIP 65426 panel of Figure \ref{fig:contrastcurves}) at the separation of the injected planet, respectively. The separation of each planet corresponds to the color of each line. The top panels correspond to the number of iterations used in our experiments, with the stopping condition marked with a dashed line (11000 iterations) and including the SNR and signal loss for 2000 iterations beyond our stopping condition. Planets injected at most separations reach SNR plateaus at similar numbers of iterations, with no strong dependence on the brightness of the injection (left and right panels). The bottom panels show the optimization resulting from significantly increasing the number of iterations, where we set our learning rate scheduler to have 20000 total iterations.}}
    \label{fig:injections_snr_sl}
\end{figure}
\begin{figure}[h!]
    \centering
    \includegraphics[width=\linewidth]{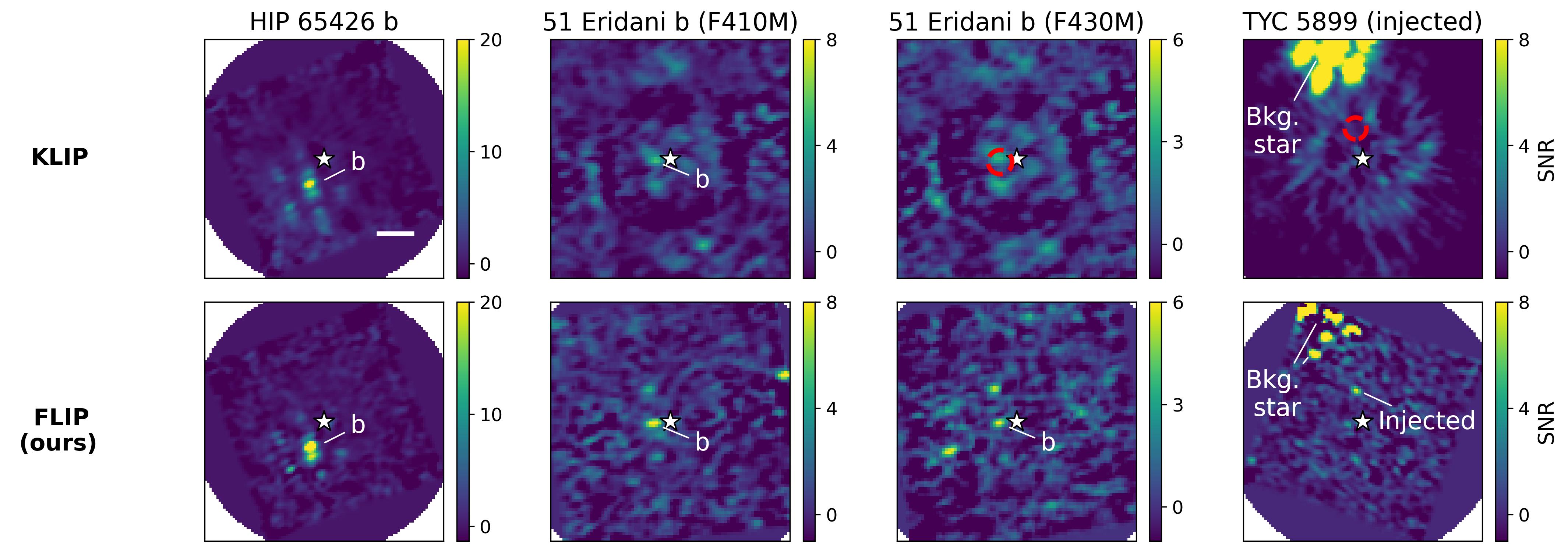}
    \caption{\textcolor{black}{\textbf{SNR maps for Figure \ref{fig:subtractions}}. In FLIP, in all cases the planet signal corresponds to the highest SNR in the scene (other than the background star for the TYC 5899 dataset). The F410M image of 51 Eridani shows an artifact on the right edge of the field of view (visible also in Figure \ref{fig:subtractions}). The F430M image of 51 Eridani also exhibits a signal to the left of the planet (visible too as a source in Figure \ref{fig:subtractions} flux maps, much fainter than the real companion); the nature of this signal is inconclusive.}}
    \label{fig:SNRmapsFig2}
\end{figure}

\begin{figure}[h!]
    \centering
    \includegraphics[width=\linewidth]{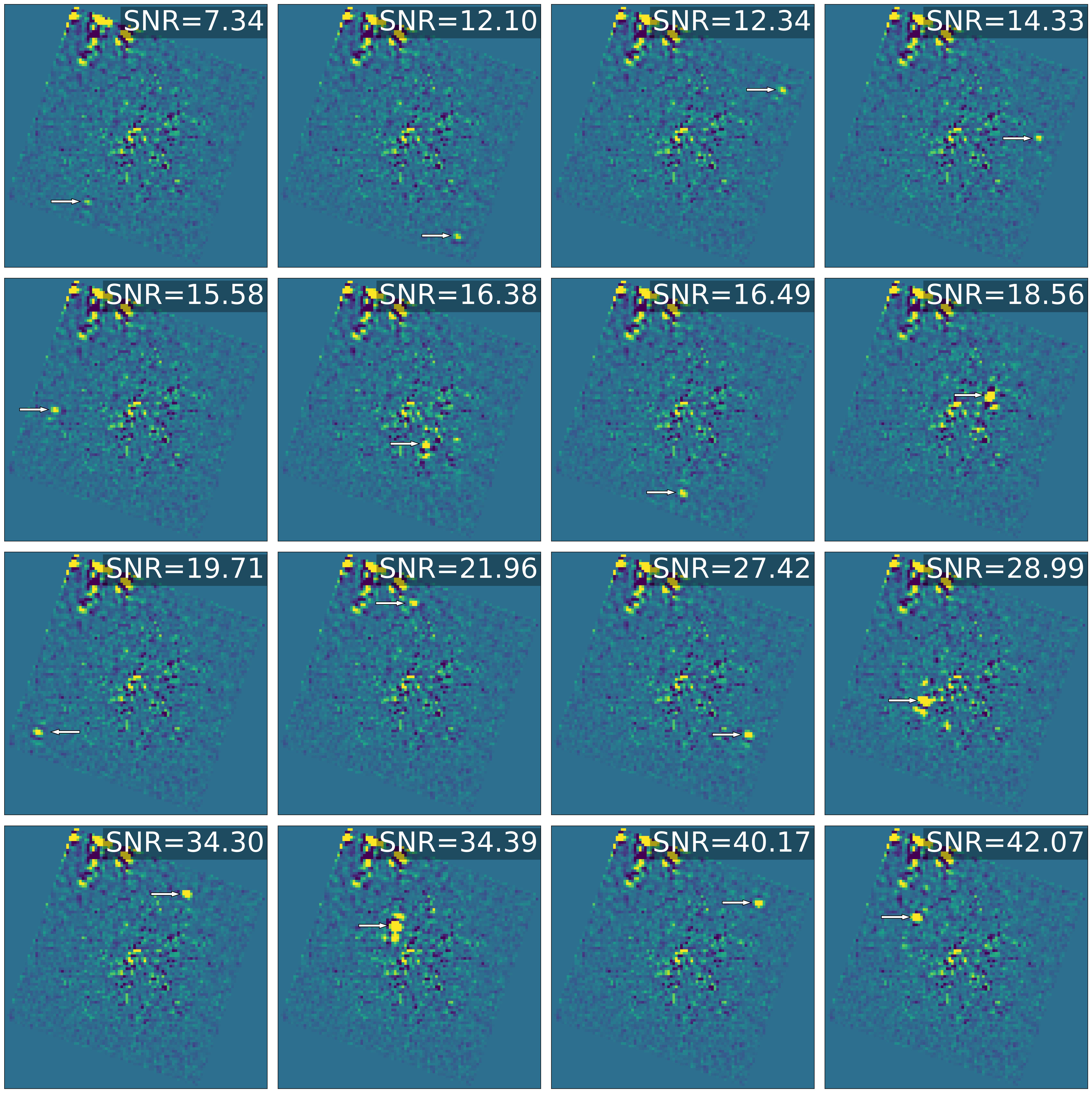}
    \includegraphics[width=\linewidth]{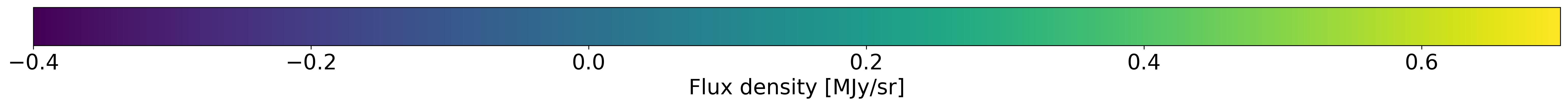}
    \caption{\textcolor{black}{\textbf{Suite of injection and recovery tests with FLIP.} We perform these on the TYC 5899 dataset. The location of the synthetic planet is annotated with an arrow. The nearby off-axis bright star is also visible in the upper left. }}
    \label{fig:ourgallery}
\end{figure}

\begin{figure}[h!]
    \centering
    \includegraphics[width=\linewidth]{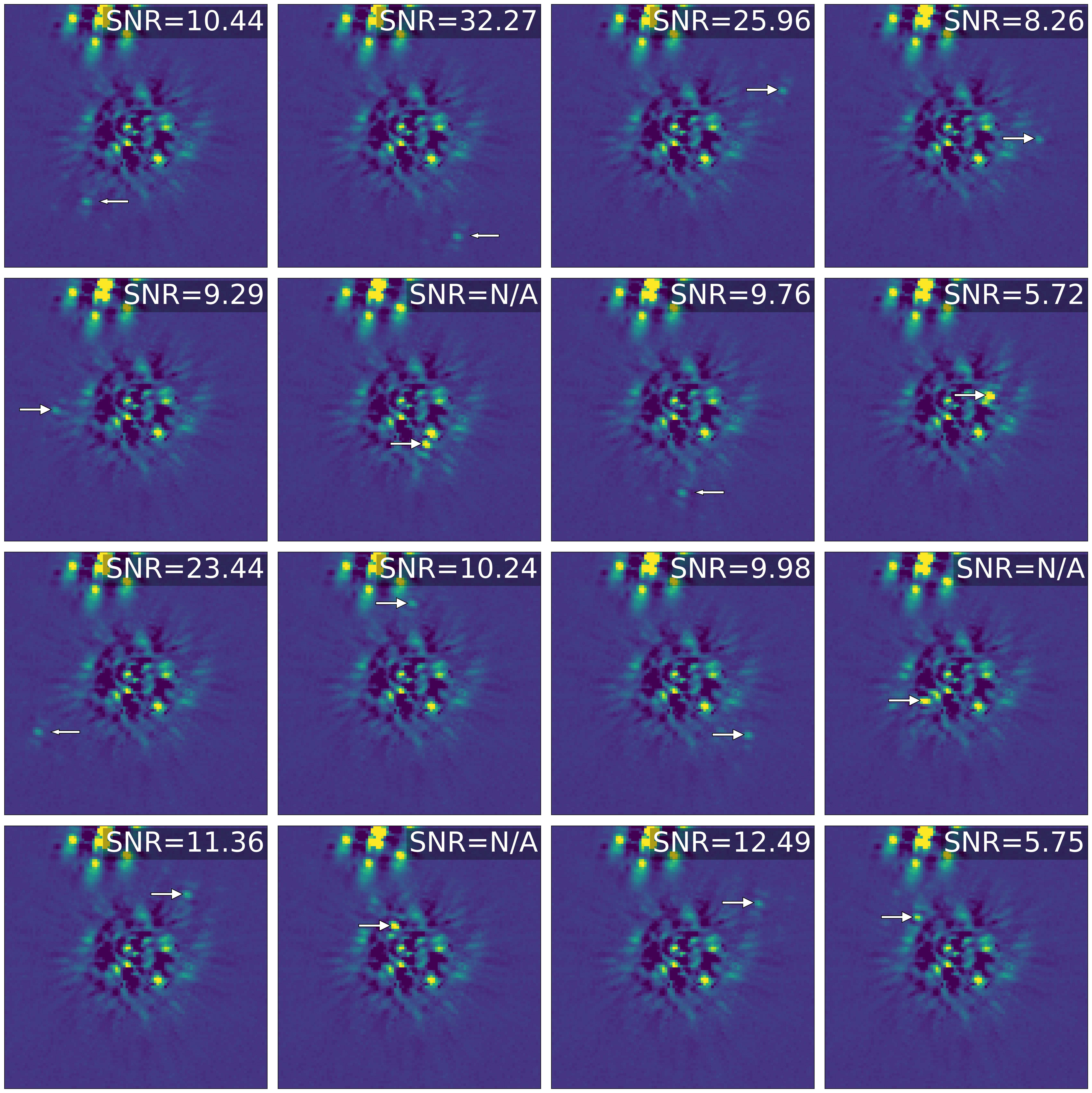}
    \includegraphics[width=\linewidth]{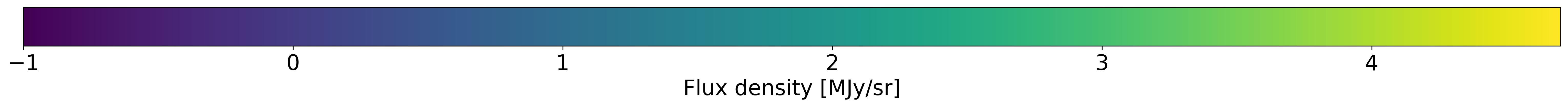}
    \caption{\textcolor{black}{\textbf{Suite of injection and recovery tests with KLIP.}Same as Figure \ref{fig:ourgallery}, but using KLIP for PSF subtraction. All injections that resulted in SNR lower than 5 are labeled as SNR=N/A.}}
    \label{fig:klipgallery}
\end{figure}

\clearpage

\captionsetup[table]{name=Extended Data Table}

\begin{table}[h]
\centering
\includegraphics[width=\linewidth]{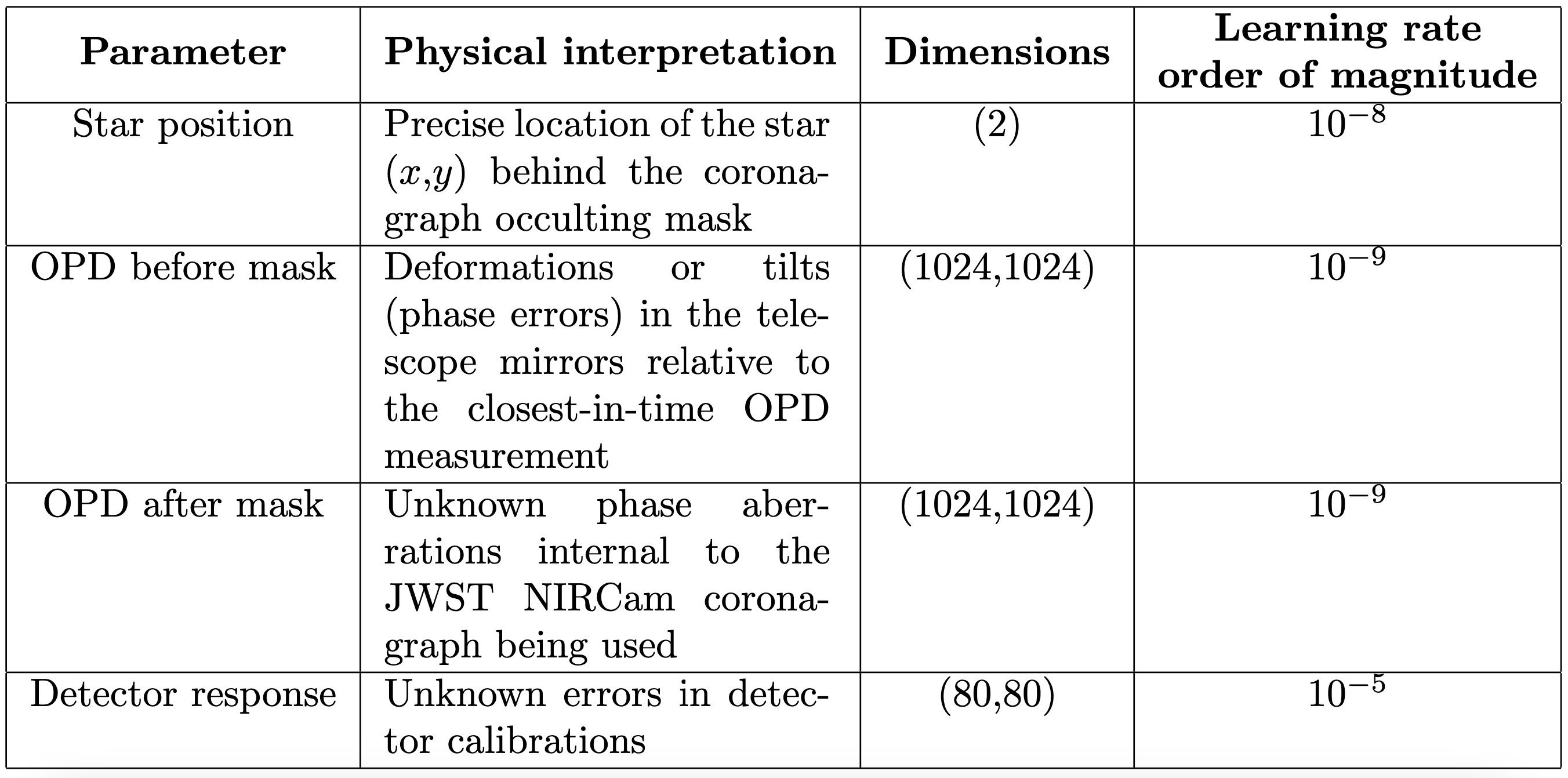}

\caption{\textbf{Summary of free parameters used in our model}. We show their physical meaning and characteristic learning rates during our optimization. The dimensionality of the detector response parameters depends on the field of view being optimized over (in our case, 80$\times$80 pixels in all cases). The flux of the star is scaled automatically to match the total flux across the image during the first sub-stage of stages 1 and 2. As mentioned in the text, while other physical mechanisms can influence the stellar PSF, we find that these four groups of parameters are sufficient to model the PSF \textcolor{black}{to the contrast levels shown in Figure \ref{fig:contrastcurves}}. For this reason, in this work the inferred parameters do not necessarily reflect the true state of the instrument, rather they produce an effective model that is flexible enough to reproduce a variety of systematics. See \textcolor{black}{Supplementary Discussion}.}
\label{tab:parametersummary}
\end{table}

\begin{table}[h!]


\includegraphics[width=\linewidth]{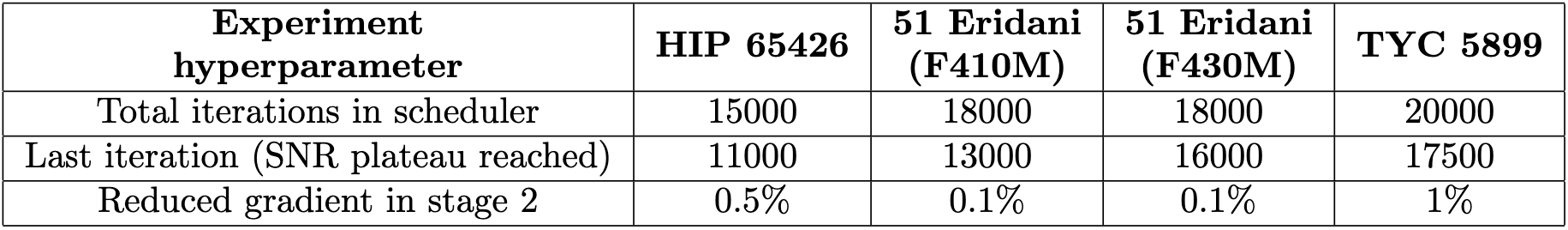}

\caption{\textcolor{black}{\textbf{FLIP hyperparameters for each experiment}. The reduced learning rate is relative to the global learning rate in stage 1.}}
\label{tab:experiment}
\end{table}














\end{bibunit}

\clearpage

\begin{bibunit}[sn-nature-custom]

\begin{center}
    \Large Imaging exoplanets below the telescope stability limit: Supplementary Information
\end{center}

\setcounter{section}{0}

\section{Supplemental Information}

\subsection{Supplementary Methods}\label{supp_methods}

\textbf{Model design implementation details.} Our physical model is initialized with the nominal optical system of \texttt{STPSF}, and we implement it in PyTorch to make it differentiable. We take the mask designs of the optical elements and the nominal wavelength and field-dependent phase aberrations.

On top of the nominal \texttt{STPSF} model, we perform a sub-pixel shift to align the center of the PSF (by default at the center of the output array) to the center of the coronagraph on the detector as measured for a given coronagraph mask and filter. We do this to perform the optimization directly on the detector coordinates, and avoid artifacts induced by interpolating and shifting the data itself. Also following \texttt{STPSF}, the PSF is convolved with a small Gaussian kernel ($\sigma=$ 0.018 arcseconds, value taken from \texttt{STPSF}; \url{https://stpsf.readthedocs.io/en/latest/jwst_detector_effects.html}) to mimic a basic charge diffusion effect. Finally, each pixel is multiplied by a pixel response function.

We parametrize the OPDs as square arrays of 1024$\times$1024 pixels, each pixel containing the OPD value in meters\textcolor{black}{, in order to have a flexible OPD model.}

Following recommendations from \texttt{STPSF}, we perform our simulations with pixel over-sampling of 2 in the image plane, and we use 21 monochromatic wavefronts for F444W observations and 9 for F410M and F430M for the best balance between numerical accuracy and evaluation speed.

\textbf{Optimization design implementation details.} \textcolor{black}{The process of parameter tuning described (number of iterations, stopping condition) is analogous to tuning KLIP hyperparameters, such as choosing the number of annuli and KL modes to construct the PSF model. In KLIP, this involves finding the parameters that best reveal a particular astrophysical signal or injecting and recovering fake planets to find the best sensitivity, and verifying that a particular source is not strongly dependent on the number of KL modes used (e.g., \supercite{sanghi2025worlds}). In FLIP, injecting fake planets and tracking their SNRs is analogous to this. In practice, we also verify that slightly tweaking these parameters does not significantly change an astrophysical source (e.g., \textcolor{black}{Extended Data Figure 7} .}
\textcolor{black}{In this work, we perform our optimization in the entire $80\times80$-pixel image. Additionally,} while our optimization can be done on \textcolor{black}{every integration separately}, we found that optimizing on the time-averaged observations yielded better contrast results for much shorter computation time. This is probably because KLIP subtractions are currently limited by static or very slowly-evolving aberrations in JWST, as opposed to other shorter-timescale aberrations \supercite{telfer2024empirical}\textcolor{black}{, so mean-combining provides a higher SNR observation, and higher SNR on the static components of the PSF in particular that need to be fit}. \textcolor{black}{On deep observations around bright stars, however, the time evolution of starlight residuals may be significant. Future work may explore optimizing on individual frames or series of frames, leveraging the temporal correlations in starlight residuals (e.g., adding regularization terms that impose similarity between frames or that formulate an explicit time evolution model).}

\textbf{PSF subtraction performance details.} To compute the SNR of each planet detection in Figure \textcolor{black}{2}, \textcolor{black}{we smooth the images ($\sigma=1$ px) and we fit a Gaussian to the core of the planet PSF at the known location of the signal. Next, we extract the peak flux of the Gaussian and divide it by the standard deviation of pixels within an annulus at the separation of the planet and with width of 0.2 arcseconds (except for HIP 65426, for which we use an annulus of width 0.5 arcseconds due to the large section of the image where the planet PSF is the dominant signal)}. \textcolor{black}{SNR is converted to a confidence level using t-statistics \supercite{mawet2014fundamental}, where we take the number of samples that determine the Student-t distribution to be the number of $3\times3$-pixel boxes (roughly the size of the resolution element) within the annular region.}

\subsection{Supplementary Discussion}\label{supp_discussion}

\subsubsection{Future improvement details}\label{methods_futuredetails}

While our experiments show significant improvements relative \textcolor{black}{to the state-of-the-art technique KLIP} on a variety of datasets, FLIP has not reached the fundamental noise limit yet. The findings in this work lead to the future avenues for improvement shown below: 

\textbf{Detector model:} Our simplified detector model is sufficient to model the majority of observations taken by JWST NIRCam. However, observations of very bright stars or very deep observations require a more sophisticated treatment of the complex detector effects present in JWST \supercite{goudfrooij2024algorithm}. This is because pixel-level defects or pipeline-induced artifacts are significantly above the noise limit and accurate modeling becomes essential. Sophisticated effective detector models or models of the data pre-processing pipeline (following \supercite{desdoigts2025amigo}) are promising to push this technique to the very high contrast regime. \textcolor{black}{Furthermore, careful treatment of detector systematics will likely be crucial to extend this framework to JWST MIRI \supercite{argyriou2023brighter}.}

\textbf{Inferred instrument parameters:} Coronagraphy is an ill-posed and degenerate problem \supercite{Xin_2024, Pope_2021}, and as such there is a wide range of solutions that may produce effectively the same PSF estimate. Furthermore, the nominal \texttt{STPSF} models for JWST NIRCam, which we use to initialize our differentiable model, are known to be imperfect. To use FLIP to retrieve realistic instrument parameters \textcolor{black}{with rigorously validated uncertainties} future calibration work is required. This could include placing strong regularization priors on some of the model parameters (e.g., \supercite{feng2024event}), jointly fitting a large number of observations or processing datasets with multiple stars in the field of view with multiple point source models. Successful calibration will enable retrieving the state of the instrument during any science observations in a physically realistic way, contributing to more accurate understanding and modeling of JWST \textcolor{black}{NIRCam}. 


\textbf{Computational cost:} For the datasets analyzed in this work, the runtime of FLIP on an NVIDIA A30 GPU varies from 15 to 35 minutes. This is very efficient for a model with $10^6$ free parameters, but exoplanet imaging routinely requires multiple runs (e.g., to calibrate total sensitivity as in Figure \textcolor{black}{3}). Addressing systematic model mismatch (improving our initial conditions) and implementing regularization on some parameters are two ways to potentially improve the speed of FLIP, perhaps close to the one minute runtime shown in monochromatic simulated data \supercite{feng2025exoplanet}.

\textbf{Scenes with bright astrophysical sources:} For simplicity, the model presented here only models the stellar PSF of the host star. As seen in Figure \textcolor{black}{2}, however, in some cases other astrophysical sources, such as bright off-axis stars, close binaries or debris disks are also present in the scenes. With no other constraints, our model ends up partially distorting these other objects. Future work on model calibration and parameter regularization could slightly reduce the flexibility of our model to avoid distorting other components in the scene. Alternatively, joint modeling of both the host star and these objects is promising for simultaneously minimizing signal distortion and characterizing the properties of these objects.

\textbf{Applicability to ground-based observations:} In a space-based instrument like JWST, the dominant source of instability for most datasets comes from drift between the reference star observation and the science observation. For this reason we obtain good results performing our optimization only once or only a handful of times per dataset (averaging all or a few subsets of the frames of the observation). However, atmospheric turbulence in ground-based observations evolves on much faster timescales, and so does the stellar PSF. 

While FLIP can be applied to ground-based observatories in principle, simply forward modeling a larger number of wavefront sensing measurements quickly becomes very computationally expensive: ground-based adaptive optics systems measure wavefront errors roughly every millisecond during every science observation, while JWST does it only once roughly every two days. Wavefront sensing data, however, should also significantly help constrain the problem, since they are taken simultaneously with the science observation, unlike JWST. 

Future work addressing this data volume challenge \textcolor{black}{(and associated sources of aberrations beyond atmospheric turbulence\supercite{Cantalloube2020,WangL2024,Levraud2024})} will be critical to extend this framework to ground-based telescopes. For example, empirical methods have been able to partially manage the high dimensionally of the problem using sparsity regularization (e.g., \supercite{bonse2025use}). \textcolor{black}{Recent work by \supercite{Kuznetsov2026} has also explored non-coronagraphic PSF reconstruction by combining physical models with deep learning models and a small set of physical parameters related to observing conditions, without relying on full adaptive optics telemetry. While the high level of PSF fidelity required for high-contrast imaging will likely benefit from explicit telemetry, frameworks like the one presented in \supercite{Kuznetsov2026} could be a promising starting point, especially when full adaptive optics telemetry is not available. These and other} approaches could be used in future iterations of FLIP to reduce the computational load on ground-based datasets.

\putbib[sn-bibliography]

\end{bibunit}

\end{document}